\documentclass{aastex62}

\usepackage{amssymb}

\usepackage{hyperref}
\usepackage{longtable}
\usepackage{gensymb}
\usepackage{lineno}
\usepackage{amsmath}
\usepackage{wasysym}
\usepackage{float}
\usepackage{afterpage}
\usepackage[version=4]{mhchem}
\usepackage{graphicx}
\usepackage{natbib}
\graphicspath{{./}}

\received{\today}
\revised{\today}
\accepted{\today}
\submitjournal{ApJ}

\shorttitle{Outer HZ weathering}
\shortauthors{Graham and Pierrehumbert}

\begin{document}

\title{Carbon cycle instability for high-$\mathrm{CO_2}$ exoplanets: Implications for habitability}

\correspondingauthor{R.J. Graham}
\email{arejaygraham@uchicago.edu}

\author[0000-0001-9289-4416]{R.J. Graham}
\affil{University of Chicago, Department of Geophysical Sciences\\
5734 S Ellis Ave.\\
Chicago, IL 60637, USA}

\author[0000-0002-5887-1197]{R.T. Pierrehumbert}
\affiliation{University of Oxford, Department of Atmospheric, Oceanic, and Planetary Physics \\
Clarendon Laboratory, Parks Rd.\\
Oxford, OX1 3PU, UK}



\begin{abstract}
Implicit in the definition of the classical circumstellar habitable zone (HZ) is the hypothesis that the carbonate-silicate cycle can maintain clement climates on exoplanets with land and surface water across a range of instellations by adjusting atmospheric $\mathrm{CO_2}$ partial pressure ($p\mathrm{CO_2}$). This hypothesis is made by analogy to the Earth system, but it is an open question whether silicate weathering can stabilize climate on planets in the outer reaches of the HZ, where instellations are lower than those received by even the Archean Earth and $\mathrm{CO_2}$ is thought likely to dominate atmospheres. Since weathering products are carried from land to ocean by the action of water, silicate weathering is intimately coupled to the hydrologic cycle, which intensifies with hotter temperatures under Earth-like conditions. Here, we use global climate model (GCM) simulations to demonstrate that the hydrologic cycle responds counterintuitively to changes in climate on planets with $\mathrm{CO_2}$-$\mathrm{H_2O}$ atmospheres at low instellations and high $p\mathrm{CO_2}$, with global evaporation and precipitation decreasing as $p\mathrm{CO_2}$ and temperatures increase at a given instellation. Within the MAC weathering formulation, weathering then decreases with increasing $p\mathrm{CO_2}$ for a range of instellations and $p\mathrm{CO_2}$ typical of the outer reaches of the HZ, resulting in an unstable carbon cycle that may lead to either runaway $\mathrm{CO_2}$ accumulation or depletion of $\mathrm{CO_2}$ to colder (possibly Snowball) conditions.  While the behavior of the system has not been completely mapped out, the results suggest that silicate weathering could fail to maintain habitable conditions in the outer reaches of the nominal HZ.
\end{abstract}



\section{Introduction}\label{sec1}

As conventionally defined \citep{Kasting93,Kopparapu:2013}, the Habitable Zone (HZ) is predicated
on habitable conditions being maintained by the joint greenhouse effect of $\mathrm{CO_2}$ and water vapor, possibly as modified by background gases such
as $\mathrm{N_2}$. The inner edge is defined by the runaway greenhouse threshold, in
which water vapor provides the dominant greenhouse effect. The outer edge is defined
by the maximum $\mathrm{CO_2}$ greenhouse effect, in which the dominant greenhouse
effect is provided by $\mathrm{CO_2}$. Actual habitability within the HZ is 
contingent on the planet having an atmosphere, and moreover having $\mathrm{CO_2}$
in the right range to permit surface liquid water. 

The conventionally defined HZ only takes into account thermodynamic and radiative
constraints on surface temperature, though it is generally an implicit assumption
that silicate weathering feedbacks adjust atmospheric $\mathrm{CO_2}$ to a range
supporting surface liquid water, supposing thermodynamic and radiative constraints
permit such a range to exist.  The implicit assumption, needed to assure
actual habitabiity, is that geochemical feedbacks keep $\mathrm{CO_2}$ from
getting too high in concentration near the inner edge, and keep $\mathrm{CO_2}$ from staying too low near the outer edge; without such a thermostat mechanism, habitability would be contingent on fortuitous fine-tuning of $\mathrm{CO_2}$ concentration.  However, there is growing recognition that
the geochemistry of the deep carbon cycle also provides constraints on $\mathrm{CO_2}$, and might fail to keep $\mathrm{CO_2}$ in the required range \citep{noack2017volcanism,foley2018carbon, foley2019habitability}. On this basis one can define a {\it geochemical HZ}, which layers constraints from the deep carbon cycle on the usual thermodynamic and radiative constraints. In addition to a carbon cycle equilibrium existing in a habitable range of $\mathrm{CO_2}$, it is required that the equilibrium be a stable equilibrium. Stability requires that weathering rates increase with increasing $\mathrm{CO_2}$. In this article, through coupled climate-weathering modeling, we exhibit some indications that the geochemical HZ may be significantly contracted relative to the conventional HZ, via destabilization of the carbon cycle equilibrium in the outer reaches of the conventional HZ.  

The estimated width of the HZ is an important input in the design of the next generation of space telescopes, some of which hope to observe and characterize the atmospheres of a handful of ``Earth analogues'' \citep{guimond2018direct,luvoir2019luvoir,HABEX_StudyReport2019,LIFE2021a}. In particular, a narrower HZ requires a larger telescope to find a given number of these planets \citep[e.g.][]{kasting2013inner}, so, in addition to potentially determining the fate of untold numbers of extraterrestrial worlds, the additional constraints defining the geochemical HZ may have practical implications for space mission design.

Intensification of the hydrologic cycle with surface temperature is one of the most robust features to emerge from studies modeling Earth's warmer future climate under the influence of higher CO$_2$ \citep{held2006robust,o2008hydrological,kundzewicz2008climate,o2012energetic,allan2020advances}, and the same basic behavior emerges under diverse simulated exoplanetary climate conditions \citep{xiong2022smaller}. Earth's global-mean precipitation rate is consistently predicted to increase approximately linearly with global-mean surface temperature at $\approx$2$\%$ per Kelvin \citep[e.g.][]{held2006robust,o2008hydrological,o2012energetic}, lagging well behind the $7\%$ that would be expected from a naive application of the Clausius-Clapeyron relation because of simultaneous slowdown in atmospheric circulation under warming conditions \citep{vecchi2006weakening}. This coupling between hydrologic cycling and surface temperature is a fundamental component of some formulations of th
e silicate weathering feedback \citep{Walker-Hays-Kasting-1981:negative}, particularly the model described in \citet{maher2014hydrologic} (MAC), where it becomes the primary mechanism by which silicate weathering is sensitive to global surface temperature, suggesting precipitation's temperature sensitivity may play a crucial role in the long-term regulation of the climates of Earth and exoplanets with carbonate-silicate cycles like Earth's \citep{graham2020thermodynamic}.

te for liquid water at a given instellation.

Given the potential importance of precipitation-climate coupling for climate stability and long-term planetary habitability, it is an underappreciated fact that the energetic cost of evaporating water into the atmosphere places a serious constraint on the maximum rate of global-mean precipitation a planet at a given instellation can sustain \citep{pierrehumbert1999subtropical,Pierrehumbert-2002:hydrologic,o2008hydrological,lehir2009snowball,Pierrehumbert:2010-book,o2012energetic}. This is readily apparent from a bulk representation of the steady-state surface energy budget:
\begin{linenomath}
\begin{align}
    S_{\rm abs} &= H_{\rm rad,sens} + L
\end{align}
\end{linenomath}
where $S_{\rm abs}$ is the absorbed instellation at the surface, $H_{\rm rad,sens}$ is the combined flux from longwave (infrared) and sensible (dry turbulent) heating/cooling of the surface, and $L$ is the latent heat flux from the surface. If we raise the concentration of CO$_2$ enough for a large amount of water vapor to accumulate in the boundary layer, longwave cooling of the surface will be suppressed and boundary layer stability will increase as the temperature contrast between the ground and the overlying air is reduced \citep{Pierrehumbert-2002:hydrologic,Pierrehumbert:2010-book,o2008hydrological}. Each of those effects acts to reduce the magnitude of term $H_{\rm rad,sens}$, while the evaporative flux (and therefore $L$) increases, which, at steady-state, is equivalent to the statement that the mean precipitation increases \citep{pierrehumbert1999subtropical}. Eventually, at high enough temperatures, the latent heat flux dominates the right side of the equation and a
pproaches the value of the absorbed instellation, such that nearly all incoming radiation goes into driving evaporation. Beyond this point, the only way to drive evaporation higher is to form an inversion layer at the surface, such that a sensible heat flux can be directed downward into the surface (i.e. turning $H_{\rm rad,sens}$ negative so that $L$ can exceed $S_{\rm abs}$) \citep{Pierrehumbert-2002:hydrologic,o2012energetic}, but the magnitude of this possible overshoot is limited \citep{o2008hydrological,Pierrehumbert:2010-book}. Thus the instellation absorbed at the surface of a planet places a fairly robust upper cap on global rates of precipitation. GCM simulations have suggested that this phenomenon may have throttled weathering rates during the hot, high-CO$_2$ aftermath of Earth's snowball events
\citep{lehir2009snowball} and low-dimensional exoplanet weathering and climate models have recently included a crude parameterization of the effect \citep{graham2020thermodynamic,coy2022diskworld}, but the implications of a maximum global precipitation rate for the stability of exoplanetary climate are still largely unexplored.

Here, for the first time, we combine global climate model (GCM) simulations with the continental weathering model introduced in \citet[][hereafter ``the MAC weathering model'']{maher2014hydrologic} and elaborated in \citet{winnick2018relationships,graham2020thermodynamic, hakim2021lithologic} to investigate carbon cycle stability in the outer reaches of the HZ. Treatment of seafloor weathering in analogous conditions is left to future work. The MAC weathering model accounts for the fact that formation of secondary minerals (i.e. clays and silica) in the weathering zone can lead to chemical equilibration that caps the concentration of weathering products in runoff, shifting the main temperature feedback in the carbonate-silicate cycle from the kinetics of silicate dissolution \citep[e.g.][hereafter ``the WHAK weathering model'']{Walker-Hays-Kasting-1981:negative} to the temperature sensitivity of hydrologic cycling. The combination of reduced top-of-atmosphere (TOA) instellati
on and elevated albedo from high CO$_2$ partial pressures ($p$CO$_2$) places strong upper bounds on the amount of global precipitation planets in low-instellation regimes can generate. This led us to suggest based on global-mean simulations in \citet{graham2020thermodynamic} that the energetic limit on precipitation might lead to warmer climates than previously expected in the outer reaches of the HZ. Full 3D GCMs conform to the energy limit on precipitation, but also capture aspects of the hydrological cycle inaccessible to energy balance modelling.  We find find that when weathering is calculated according to the MAC formulation using GCM-based precipitation and temperature fields, the carbonate-silicate cycle transitions from a negative, stabilizing feedback for planets at high instellation and low $p$CO$_2$ to a positive, destabilizing feedback for planets with energetically-limited hydrology due to low instellation and high $p$CO$_2$.

The resulting climate-carbon equilibrium in the latter regime, in which volcanic outgassing is balanced by sinks due to silicate weathering is unstable. If the equilibrium is displaced on the low $\mathrm{CO_2}$ side, the $\mathrm{CO_2}$ will continue to decrease until it finds a new equilibrium with low $\mathrm{CO_2}$, resulting in a colder, potentially Snowball, state. If displaced towards the high $\mathrm{CO_2}$ side, $\mathrm{CO_2}$ will continue to accumulate in the atmosphere, resulting in either an uninhabitably hot state or a state with a temperate liquid $\mathrm{CO_2}$ ocean, depending on instellation.  In contrast, when weathering is calculated according to WHAK (which exhibits a greater degree of direct temperature dependence), the carbon cycle uniformly displays negative feedback behavior in the high $\mathrm{CO_2}$ outer portion of the HZ we have probed as well as in the inner low $\mathrm{CO_2}$ . 

In essence, our calculations have identified a forbidden zone of the climate/carbon equilibrium extending to at least the range of $\mathrm{CO_2}$ between approximately 1 and 4 bars, and instellation between 675 and 1000 $\mathrm{W/m^2}$, within which the climate/carbon equilibrium is unstable.  However, because of limitations in our modelling framework, we have not identified the precise boundaries of the forbidden zone, and in particular have not been able to probe instellations all the way to the outer edge of the conventional HZ.  The instability in the forbidden zone implies a novel form of hysteresis in the climate-carbon system. Climates in the forbidden zone will be attracted to low or high $\mathrm{CO_2}$ outside the forbidden zone, but we are not currently able to precisely identify these attractors, though we offer some speculations about what they might be. While our results are not definitive, they have uncovered  some very novel behavior in the outer reaches of 
the conventional HZ, which may restrict the geochemical HZ to a smaller range of orbital distances than the conventional HZ. 

In Section \ref{sec:ch4_methods} we describe the models and methods we applied, in Section \ref{sec:ch4_results} we describe the results of our simulations,in Section \ref{sec:ch4_discussion} we discuss the implications of these results, in Section \ref{subsec:ch4_escape_routes} we discuss some important caveats concerning our results, and in Section \ref{sec:ch4_conclusion} we summarize our main findings. 

\section{Methods}\label{sec:ch4_methods}

\subsection{Climate model}
\begin{figure*}[htb!]
    \centering
    \makebox[\textwidth][c]{\includegraphics[width = 300pt,keepaspectratio]{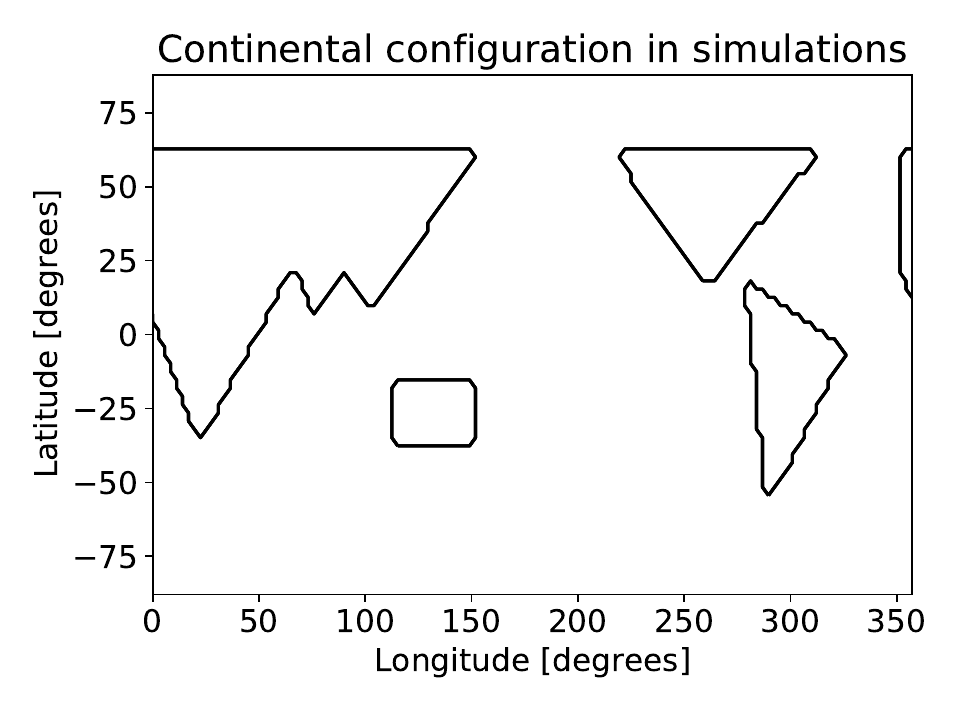}}
    \caption{\textbf{The continental configuration used in the presented simulations.}}
\label{fig:continent_config}
\end{figure*}

To simulate planetary climate, we use the open-source Isca general circulation model (GCM) framework \citep{vallis2018isca}, which solves the hydrostatic pressure-coordinate primitive equations on a sphere using a pseudospectral dynamical core. The model has been used in a wide variety of planetary climate contexts \citep[e.g.][]{penn2018atmospheric,thomson2019effects,thomson2019hierarchical, yang2019simulations}. We present fourteen GCM simulations, with eleven in high-$p$CO$_2$, low-instellation model configurations and three with more Earth-like configurations (discussed further below; see also Table \ref{tab:ch4_gcm}). All simulations have a T42 horizontal resolution (64 latitudes, 128 longitudes) and 40 vertical sigma-pressure layers. We employ built-in Isca features, notably a simple Betts-Miller convective relaxation scheme \citep{betts1993betts} and a bucket hydrology configuration governing evaporation \citep{vallis2018isca}, modeled after the treatment of soil in \c
itet{Manabe-1969:climate}. Boundary layer fluxes are treated with Isca's default Monin-Obukov scheme \citep{vallis2018isca} as in \citet{frierson2006gray}. The land configuration in the simulations we present consists of a simplified, topography-free polygonal representation of modern-day continents (Fig. \ref{fig:continent_config}), but qualitatively similar results emerged from a smaller set of simulations with a single large, equatorial continent (not shown), suggesting the basic physical phenomena in play are insensitive to continental configuration. A full description of the Isca model can be found in \citep{vallis2018isca}, and the model outputs and post-processing scripts used for this article are available for download \citep{graham_data_2024}. Planetary surfaces are given a uniform albedo of 0.05, comparable to that of Earth's seawater at the insolation-weighted global-mean zenith angle of $\approx$48.19$\degree$ \citep{li2006ocean,cronin2014choice}. This is a conservative c
 hoice, since applying a higher albedo to land would reduce the amount of sunlight absorbed at the planetary surface and lower the energetic limit on precipitation. We use Isca's slab ocean configuration with a reduced, 10-meter mixed layer depth to accelerate convergence of the simulations to energetic steady-state. We discuss the potential limitations of a slab ocean configuration in Section \ref{subsec:oht}. Obliquity and eccentricity are set to zero, so there is no seasonal cycle. The day length in all simulations is 24 hours, so our results are specific to rapidly rotating planets like Earth. Isca's current implementation is cloud-free, a significant simplification, but this study already explores a number of highly novel aspects at the intersection of hydrology and weathering, so it was our judgement that it is at this point best to stick to the relatively robust clear-sky physics without layering on additional uncertainties arising from the many different ways cloud feedbacks 
 can behave. Certainly, exploration of the extent to which the behavior we reveal survives the addition of cloud effects is a prime target for future research.  It is worth noting that under high CO$_2$ conditions, low-lying high-albedo water clouds are predicted to decrease substantially in extent due to the inhibition of cloud-top radiative cooling in a highly opaque atmosphere \citep{schneider2019possible,goldblatt2021earth}, so the approximation may be less grave under those conditions; even if this effect carries over to the higher $\mathrm{CO_2}$ regime in the present study, the effect of higher altitude water clouds could be significant.  $\mathrm{CO_2}$ clouds can form for planets quite close to the outer edge of the HZ, but we only present simulations where CO$_2$ is subsaturated throughout the atmosphere, so although CO$_2$ clouds may be relevant in similar climates with slightly higher $p$CO$_2$ or lower temperatures \citep[e.g.][]{kitzmann2017clouds}, their neglect has no
  impact on our results. The model also excludes land or sea ice, so the ice-albedo feedback \citep[e.g.][]{Sellers-1969:ae} is not in play, but this is not a serious issue for the warm climates we are examining, all of which sport little or no surface area with temperatures below freezing (as described in Section \ref{subsec:ch4_temperature_results}). Further, the ice-albedo feedback is weakened significantly under thick CO$_2$ atmospheres, which reduce the impact of changes to the surface albedo on planetary energy budgets \citep{von2013dependence}.

Radiative transfer calculations coupled to Isca's dynamics are carried out with the \textsc{socrates} code \citep{edwards1996studies}, using the correlated-k method to solve the plane-parallel, two-stream approximated radiative transfer equation with scattering for atmospheres irradiated by a solar (G-star) spectrum. In high-instellation, low-$p$CO$_2$ simulations, we used the standard, validated spectral files used by the UK Met Office in the latest iteration of its Unified Model to simulate Earth's climate \citep{walters2019met}. For high $p$CO$_2$ model configurations, we used a spectral file from NASA GISS's ROCKE-3D \citep{way2017resolving} database and created for GCM modeling of the ancient Martian atmosphere, similar (though not identical) in construction to that used in \citet{guzewich20213d} (Eric T. Wolf, personal communication). The file is valid for atmospheres up to 10 bars with $>90\%$ CO$_2$ and temperatures up to 400 K, with some loss of accuracy above 310 K 
(though we compared TOA fluxes from this model at high temperatures with those generated using a high resolution 318 band spectral file used in \citet{graham2022co2} and found the output from the GCM stayed within 5 W m$^{-2}$ of the higher resolution calculations). Rayleigh scattering is interactively calculated with coefficients included in \textsc{socrates} for the relevant gases. For CO$_2$, opacity coefficients were tabulated and derived from the HITRAN database, making use of line-by-line and collision-induced absorption coefficients, along with sub-Lorentzian line-broadening and self-broadening  \citep{perrin1989temperature,baranov2004infrared,HITRAN2016}, and similarly for H$_2$O, which included line-by-line and  collision-induced absorption coefficients and the H$_2$O MT-CKD continuum, along with broadening coefficients that are unfortunately calculated with respect to Earth air \citep{mlawer2012development,HITRAN2016}, which may somewhat underestimate water opacities at hig
 h temperatures \citep{Pierrehumbert:2010-book}. We note that the CO$_2$ continuum spectrum is uncertain at high temperatures and pressures, which introduces a potentially significant source of error into our calculations \citep[e.g.][]{Halevy09,wordsworth2010infrared}.

\begin{table*}[htb!]
\centering
\begin{tabular}{rcccccc}
\hline
&\vline&\textbf{$\mathbf{\mathit{S}}$ [W m$^{-2}$]:}&&&&\\
&\vline&675 &750& 800 & 1000 & 1250\\
\hline
\textbf{$\mathbf{\mathit{p}}$CO$_{2}$ [bars]:}&\vline&&&&&\\
\textcolor{green}{2$\times10^{-4}$}&\vline&&&&&\checked\\
\textcolor{green}{3$\times10^{-4}$}&\vline&&&&&\checked\\
\textcolor{green}{4$\times10^{-4}$}&\vline&&&&&\checked\\
\cline{2-7}
1             &\vline &&\checked&\checked&\checked&\\
2             &\vline &\checked&\checked&\checked&\checked&\\
3             &\vline &\checked&\checked&\checked&&\\
4             &\vline &\checked&&&&\\
\hline
\end{tabular}
\caption{Table with checkmarks indicating the combinations of substellar TOA instellation ($S$; columns) and CO$_2$ partial pressure ($p$CO$_2$; rows) used in the simulations presented in this study. The simulations with $p$CO$_2<$ 1 bar (marked green) are supplemented by 1 bar of N$_2$, while those with $p$CO$_2>$1 bar have no atmospheric N$_2$.}
\label{tab:ch4_gcm}
\end{table*}

With one exception, each simulation was run until the TOA and surface energy fluxes were balanced to within $<$1$\%$ and then for at least a year further, with stated results coming from averages over the final year of data for each simulation. The three low $p$CO$_2$ simulations have N$_2$-dominated atmospheres, surface pressures of 1 bar, and CO$_2$ concentrations of 200, 300, and 400 parts per million by volume (ppmv), all irradiated by TOA instellation of $S=1250$ W m$^{-2}$.  The eleven high $p$CO$_2$ simulations have TOA  substellar shortwave instellations of  $S=675$, 750, 800 or 1000 W m$^{-2}$ and CO$_2$ partial pressures of one, two, three, or four bars (see Table \ref{tab:ch4_gcm}). All simulations have atmospheric H$_2$O content determined by Isca's hydrologic cycle. The simulation with three bars of CO$_2$ irradiated by $S=800$ W m$^{-2}$ is the exception to the statement about equilibration of fluxes to within $<1\%$, as one of its grid cells exceeded the model'
s temperature limit of 350 K during spin-up while TOA fluxes were still 1.9$\%$ out of balance. Thus results from that simulation are averaged over a period during which the planetary surface was still gradually heating, with a 2$\%$ TOA flux imbalance, instead of $<1\%$ like the others. Based on the behavior of the other simulations, the model would have warmed another Kelvin or two in the mean if it had been able to run to equilibrium, a small error unlikely to have any impact on the qualitative trends demonstrated here. 

\afterpage{%
\begin{longtable}[c]{|p{2.3cm}|p{2.7cm}|p{4.8 cm}|p{2.1cm}|}
\hline
\textbf{Parameter} & \textbf{Units} & \textbf{Definition} & \textbf{Fiducial Value} \\
\hline
$\gamma_{\rm Earth}$&$\textbf{--}$ & Earth's land & 0.3\\
&&fraction&\\
\hline
$a_g$ & $\textbf{--}$ & Surface albedo & 0.05\\
\hline
$R_{\rm planet}$ & meters (m) & Planetary radius & 6.37$\times10^6$\\
\hline
$W_{\rm ref}$ & moles per year & Reference weathering rate &  3.4$\times10^{12}$\\
&  & equal to recent Earth& \\
&(mol yr$^{-1}$)&  outgassing estimate&\citep{coogan2020average}\\
\hline
$T_{\rm ref}$& Kelvin (K) & Reference global-& 288\\
&&avg. temperature &\\
\hline
$p$CO$_{\rm 2,ref}$& bar & Reference &280$\times10^{-6}$\\
&&\ce{CO2} partial pressure&\\
\hline
$\Lambda$&variable&Thermodynamic &1.4$\times10^{-3}$\\
&&coefficient for $C_{eq}$&\\
\hline
$n$&\textbf{--}&Thermodynamic $p$\ce{CO2}&0.316\\
&&dependence&\\
\hline
$\alpha$* & $\textbf{--}$ & $L\phi\rho_{sf}AX_r\mu$& 3.39$\times 10^5$\\
&&(see Section \ref{subsec:ch4_weathering_models} and below)&\\
\hline
$k_{\rm eff,ref}$*&mol m$^{-2}$ yr$^{-1}$&Reference rate constant&8.7$\times10^{-6}$\\
\hline
$\beta$ & $\textbf{--}$ & Kinetic weathering & 0.2\\
& &$p$\ce{CO2} dependence &\citep{rimstidt2012systematic}\\
\hline
$T_{\rm e}$ & Kelvin & Kinetic weathering & 11.1\\
& &temperature dependence  &\citep{Berner:1994p3295}\\
\hline
\caption{Weathering parameters used in this study. This table lists parameters used in our calculations, their units, their definitions, and the default values they take. A single asterisk (*) means the default parameter value was drawn from Table S1 of the supplement to \citet{maher2014hydrologic}. For default parameters drawn from other sources, the citation is given in the ``Value'' column.
\label{tab:ch4_weathering_values}
}
\end{longtable}
}
\subsection{Weathering models}\label{subsec:ch4_weathering_models}
For both the MAC and the WHAK formulation, we calculate a weathering flux ($F_{\rm sil}$ [moles m$^{-2}$ yr$^{-1}$]) in any grid cell that 1.) has land, 2.) has a surface temperature above the triple point of water, 273.15 K, and 3.) has a local precipitation flux greater than zero. Here, we focus exclusively on continental silicate weathering, ignoring the potential contribution of analogous reactions on the seafloor \citep[e.g.][]{coogan2013evidence} due to lack of clarity about the strength of this feedback, but we return to the issue of seafloor weathering in Section \ref{subsec:ch4_escape_routes}. A global weathering rate ($W_{\rm tot}$ [moles yr$^{-1}$]) is calculated by multiplying each cell's weathering flux by the surface area ($dA$) of the grid cell and then adding them all together i.e.
\begin{linenomath}
\begin{align}
    W_{\rm tot}=\sum_\mathrm{lat,lon}F_{\rm sil}dA
\end{align}
\end{linenomath}
rendering the CO$_2$ consumption rate for a planet with the assumed weathering properties and background climate. 

For silicate weathering to serve as a stabilizing negative feedback on planetary climate, it must act to maintain the climate at a set-point determined by the properties of the silicates being weathered, the orbital and surface properties of the planet, and the planet's CO$_2$ outgassing rate. This requires global weathering to accelerate with increases to $p$CO$_2$ and surface temperature. Under such conditions, if a planet has its climate perturbed in a way that makes its weathering rate fall below its CO$_2$ outgassing rate, CO$_2$ will be consumed more slowly than it is added to the atmosphere, leading to net CO$_2$ growth and surface warming until the global weathering rate has accelerated to the point that it is equal to outgassing, whence the atmospheric CO$_2$ content will stop evolving. In the opposite scenario, where global weathering decelerates in response to increases in $p$CO$_2$ and surface temperature, the process becomes a destabilizing feedback. In this case
, the carbon/climate equilibrium where weathering matches outgassing is unstable.  If $p\mathrm{CO_2}$ is displaced to the high side of the equilibrium, weathering becomes less than outgassing, $\mathrm{CO_2}$ would accumulate, and temperature would increase until something occurs to arrest the process (e.g. the planetary interior running out of $\mathrm{CO_2}$). If $p\mathrm{CO_2}$ is displaced to the low side of the equilibrium, weathering will exceed outgassing and $p\mathrm{CO_2}$ will continue to decrease until a stable low $p\mathrm{CO_2}$ climate (possibly a Snowball) is reached. 

Thus the stability of a carbon cycle at a given instellation can be determined by calculating how the weathering rate changes with $p$CO$_2$ (the so-called ``weathering curve'' \citep{penman2020silicate}): if weathering rate is positively correlated with $p$CO$_2$, it acts as a stabilizing negative feedback, but if it is inversely correlated with $p$CO$_2$ it acts as a destabilizing positive feedback. So, rather than carry out very long asynchronously-coupled simulations evolving atmospheric $p$CO$_2$ according to the balance between an assumed outgassing flux and a calculated weathering flux, we take the much simpler approach of doing a handful of simulations at a variety of $p$CO$_2$ and diagnosing the global weathering rates from these ``snapshots.'' Simply determining the sign of the slope of the weathering curve at a given instellation is sufficient for determining the stability of the carbon cycle. Under stabilizing conditions, imposing an outgassing rate would inexorab
ly drive the $p$CO$_2$ to the level that generates a weathering rate equal to the assumed outgassing rate. Under destabilizing conditions, the final outcome would be determined by the initial conditions, i.e. whether the planet started off with outgassing greater than or smaller than its initial weathering rate.

In the next subsections, we describe how the two weathering models we deploy (MAC and WHAK) calculate $F_{\rm sil}$ as a function of local climate. 

\subsubsection{WHAK}
The formulation of weathering developed in \cite{Walker-Hays-Kasting-1981:negative} (WHAK) is the basis for the calculations in most previous exoplanet weathering/climate studies, including the few that have employed 3D GCMs \citep[e.g.][]{edson2012carbonate,paradise2017,jansen2019climates,paradisehabitable}. We carry out calculations using the WHAK weathering formulation to compare with results from the MAC model, which is more closely tied to the underlying weathering chemistry than WHAK. WHAK weathering implicitly assumes silicate weathering rates are limited by the kinetics of silicate dissolution, which produces an exponential dependence on temperature. WHAK also includes a power-law dependence on CO$_2$ partial pressure ($p$CO$_2$).  We represent this as: 
\begin{linenomath}
\begin{align}\label{eqn:ch4_whak}
    F_\mathrm{sil} = \frac{W_{\rm ref}}{\gamma_{\rm Earth}\times4\pi R_{\rm Earth}^2}\exp{(\frac{T-T_{\rm ref}}{T_\mathrm{e}})}(\frac{\text{pCO}_2}{\text{pCO}_{2,\rm ref}})^\beta
\end{align}
\end{linenomath}
where $F_\mathrm{sil}$ [mol m$^{-2}$ yr $^{-1}$] is the weathering flux from a grid cell, i.e. the number of divalent cations (which react with oceanic carbon to form carbonate minerals, ultimately removing CO$_2$ from the atmosphere) delivered to the ocean per unit time per unit land area in a given grid cell; $W_{\rm ref}$ is a reference global weathering rate (assumed equal to an estimate of Earth's modern-day CO$_2$ outgassing, see Table \ref{tab:ch4_weathering_values}); $\gamma_{\rm Earth}\times4\pi R_{\rm Earth}^2$ is the approximate land area of the Earth, calculated by multiplying its land fraction ($\gamma_{\rm Earth}$) by its surface area ($4\pi R_{\rm Earth}^2$), necessary for translating $W_{\rm ref}$ from a global weathering rate into a weathering flux per unit land area; $T$ [K] is the local surface temperature in a grid box; $T_\mathrm{e}$ is the e-folding temperature for the weathering reaction; $p$CO$_2$ [bar] is the CO$_2$ partial pressure; and $\beta$ is th
e power-law dependence on $p$CO$_2$. The values of $T_e$ and $\beta$, which determine the sensitivity of the reaction rate to changes in temperature and $p$CO$_2$, vary considerably for different silicate minerals. We choose the default values listed in Table \ref{tab:ch4_weathering_values} based on results from laboratory silicate dissolution experiments \citep[e.g.][]{schott1985dissolution, brady1991effect,knauss1993diopside,oxburgh1994mechanism,welch1996feldspar,chen1998diopside,weissbart2000wollastonite,oelkers2001experimental,palandri2004compilation,carroll2005dependence,golubev2005experimental,bandstra2008data,brantley2008kinetics}. In this formulation of WHAK, we ignore the runoff-dependence assumed in the original formulation \citep{Walker-Hays-Kasting-1981:negative}, as it has been demonstrated not to substantially impact global weathering rates, which are dominated by the exponential temperature dependence and power-law CO$_2$ dependence \citep{abbot12-weathering}. Further,
  truly kinetically-limited weathering would not be impacted by changes to runoff, since the weathering rate would instead be determined by the rate of production of cations through silicate dissolution. Absent an accompanying change to temperature or $p$CO$_2$, any change to runoff above zero would simply change the dilution of weathering products without ultimately altering their rate of production or delivery to the ocean.

\subsubsection{MAC}\label{subsubsec:MAC}
The other weathering model we apply in our simulations is modified from \citep{graham2020thermodynamic}, which applied a global-mean version of the MAC weathering model \citep{maher2014hydrologic, winnick2018relationships} to calculate global weathering fluxes from models of Earth-like exoplanets. The MAC model accounts for the impact of clay formation in the weathering zone, which sets a maximum concentration on weathering products in runoff, drastically increasing the importance of hydrology for determining weathering fluxes when water moves through the weathering zone at a rate that allows for weathering products to reach the maximum chemically-equilibrated concentration. In this paper, rather than a global-mean formulation, we apply the weathering model from \cite{graham2020thermodynamic} locally in each land grid cell, with the weathering flux in a given cell determined by the parameterized silicate properties, local H$_2$O precipitation flux, surface temperature, and $p
$CO$_2$:
\begin{linenomath}
\begin{align}\label{eqn:ch4_mac}
    F_{\rm sil} &= \frac{\alpha}{[k_{\rm eff}]^{-1}+mAt_{\rm s}+\alpha[qC_{\rm eq}]^{-1}},
\end{align}
\end{linenomath}
where $F_{\rm sil}$ [mol m$^{-2}$ yr$^{-1}$] is the weathering flux from a given cell as above; $\alpha$ is a parameter that captures the effects of various weathering zone properties like characteristic water flow length scale, porosity, ratio of mineral mass to fluid volume, and the mass fraction of minerals in the weathering zone that are weatherable (see \citet{graham2020thermodynamic} for a full explanation); $k_{\rm eff}=k_{\rm eff, ref}\exp{\left(\frac{T_{\rm surf}-T_{\rm surf,ref}}{T_e}\right)}\left(\frac{p\text{CO}_2}{p\text{CO}_{\rm 2,ref}}\right)^\beta$ [mol m$^{-2}$ yr$^{-1}$] is the effective kinetic weathering rate, i.e. the weathering rate in the absence of chemical equilibration with clay formation \citep{Walker-Hays-Kasting-1981:negative}; $m$ [kg mol$^{-1}$] is the average molar mass of minerals being weathered; $A$ [m$^{2}$ kg$^{-1}$] is the average specific surface area of the minerals being weathered; $t_s$ [yr] is the mean age of the material being weath
ered; $q$ [m yr$^{-1}$] is the flux of water through the grid cell, given by the precipitation in that grid cell output by the GCM; and $C_{\rm eq}=\Lambda(\text{\textit{p}CO}_2)^n$ [mol m$^{-3}$] is the maximum concentration of divalent cations in the water passing through weathering zones, as determined by chemical equilibrium between dissolving silicates and the secondary minerals (clays) that form from their products. The default values of all constants are given in Table \ref{tab:ch4_weathering_values}. 

By equating local precipitation and $q$, we are assuming that any rain that falls on land drives weathering that delivers solutes to the ocean, i.e. that all precipitation is converted to runoff. Thus these calculations provide an upper limit on weathering fluxes and the efficiency of MAC weathering as a climate feedback with a given set of parameters, as smaller rates of runoff to the ocean would reduce weathering rates. On modern Earth, 20-26$\%$ of precipitation falling on land may be converted to runoff \citep{ghiggi2019grun,graham2020thermodynamic,coy2022diskworld}, but this value is likely to be heavily dependent on factors like topography, surface temperature, and background atmosphere. As long as runoff is largely monotonic with precipitation, as seems likely \citep{lehir2009snowball,ghiggi2019grun} (though it is unclear whether this must be the case \citep{o2012energetic}), the qualitative behavior we describe should hold. Smaller rates of conversion from rain to run
off exacerbate the severity of the mechanism explored here by allowing the energetically-limited regime to be accessed under smaller background CO$_2$ outgassing rates. 

\section{Results}\label{sec:ch4_results}
\subsection{Surface temperature}\label{subsec:ch4_temperature_results}
The mean climate states and CO$_2$ equilibrium climate sensitivities (ECS) of the simulations are in line with previously published estimates.

\begin{figure*}[htb!]
    \centering
    \makebox[\textwidth][c]{\includegraphics[width = \textwidth,keepaspectratio]{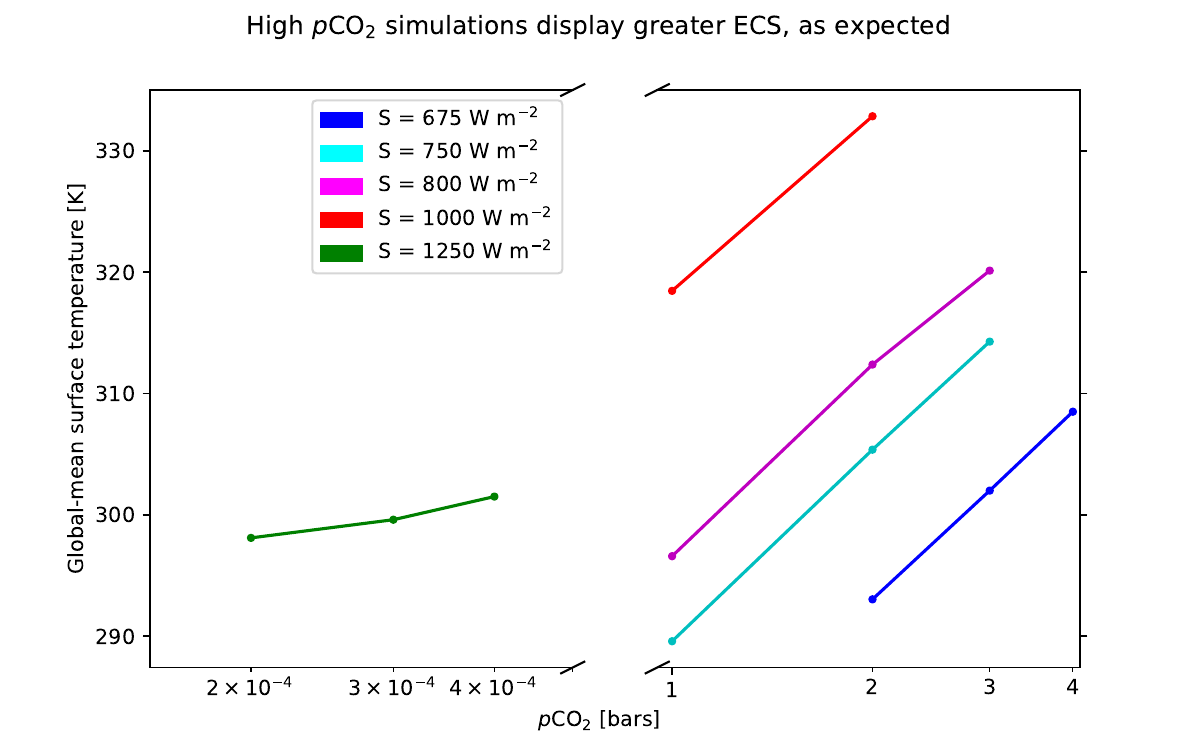}}
    \caption{\textbf{Global-mean surface temperature as a function of $p$CO$_2$ at various instellations $S$}. The low-$p$CO$_2$, high-instellation simulations display an ECS of 3.4 K, close to estimates of the modern Earth's (1.5 to 4.5 K). The high-$p$CO$_2$, low-instellation simulations all display ECS values of around 15 K due to CO$_2$'s self-broadening effect. Blue corresponds to $S=675$ W m$^{-2}$, cyan to 750 W m$^{-2}$, magenta to 800 W m$^{-2}$, red to 1000 W m$^{-2}$, and green to 1250 W m$^{-2}$.}
    \label{fig:ch4_ECS}
\end{figure*}

The ECS is the steady-state change in global-mean surface temperature from a doubling of atmospheric CO$_2$, commonly used in studies of Earth's climate \citep{knutti2017beyond,romps2020climate,goodwin2021probabilistic}. The modern Earth's ECS is generally estimated to fall between 1.5 K and 4.5 K per doubling \citep[e.g.][]{knutti2017beyond}. This is consistent with the global-mean temperature behavior of our low-$p$CO$_2$ simulations, which warmed from 298.1 to 301.5 K as CO$_2$ was doubled from 200 ppmv to 400 ppmv (see green line and dots in Fig. \ref{fig:ch4_ECS}), indicating a reasonable ECS of 3.4 K. These simulations are substantially hotter than the modern Earth despite receiving 8$\%$ less TOA instellation because they lack clouds and the surface has a uniform, dark albedo of 0.05, resulting in planetary albedo of 0.1 due to N$_2$'s modest Rayleigh scattering effect, in contrast to Earth's cloud- and ice-maintained albedo of 0.29 \citep{stephens2015albedo}. None of 
the low-$p$CO$_2$ simulations displays temperatures below freezing on any landmass, and in each case the fraction of planetary surface area below freezing at the poles is small (4.2$\%$, 2.5$\%$, and 1.7$\%$ for the 200 ppmv, 300 ppmv, and 400 ppmv simulations respectively).

In climate simulations with very high CO$_2$, ECS is known to increase as a function of $p$CO$_2$ due to the increasing importance of self-broadening and activation of spectral regions that are minor at lower pressures \citep[e.g.][]{Halevy09,Pierrehumbert:2010-book, russell2013fast,wordsworth2013water,ramirez2014can,wolf2018evaluating,graham2021high}. Our high $p$CO$_2$ simulations reflect this behavior, with all displaying an ECS of $\approx$15 K (see blue, cyan, magenta, and red lines with dots in Fig. \ref{fig:ch4_ECS}). This is consistent with the behavior of GCM simulations of early Earth atmospheres with instellations slightly larger and $p$CO$_2$ levels slightly lower than those presented here \citep{wolf2018evaluating}, as well as with ECS values calculated using a polynomial fit \citep{kadoya2019outer} to radiative-convective column model output \citep{Kopparapu:2013} with 1 bar N$_2$, saturated H$_2$O, and up to 10 bars CO$_2$ \citep{graham2021high}. None of the pr
esented high-$p$CO$_2$ simulations has temperatures below freezing on any landmass, and only the 1 bar, $S=750$ W m$^{-2}$ has any planetary surface area below freezing (5.2$\%$, at the poles). Finally, as expected, with a given $p$CO$_2$, increased instellation leads to higher surface temperatures. 

\begin{figure*}[htb!]
    \centering
    \makebox[\textwidth][c]{\includegraphics[width = \textwidth,keepaspectratio]{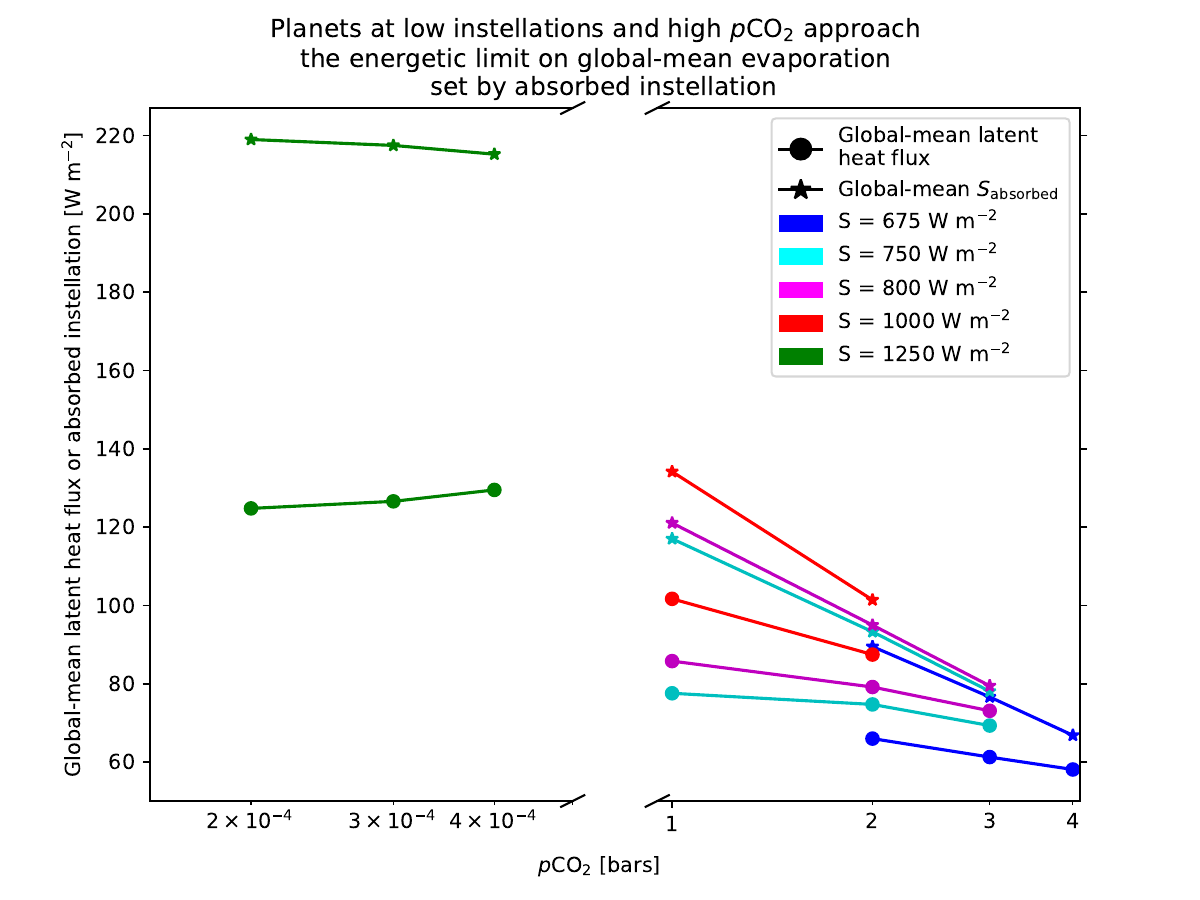}}
    \caption{\textbf{Global-mean absorbed instellation and global-mean upward latent heat flux as functions of $p$CO$_2$ at various $S$} For low-$p$CO$_2$, high-instellation simulations, latent heat flux (circle markers) increases with $p$CO$_2$, but remains far below absorbed instellation (star markers). For high-$p$CO$_2$, low-instellation simulations, a large majority of their absorbed instellation goes into the latent heat of evaporation. Colors are as in Fig. \ref{fig:ch4_ECS}.}
    \label{fig:ch4_energetics}
\end{figure*}

\subsection{Energy budgets and precipitation}
\subsubsection{Low $p$CO$_2$, high instellation}
\begin{figure*}[htb!]
    \centering
    \makebox[\textwidth-50pt][c]{\includegraphics[width = \textwidth-50pt,keepaspectratio]{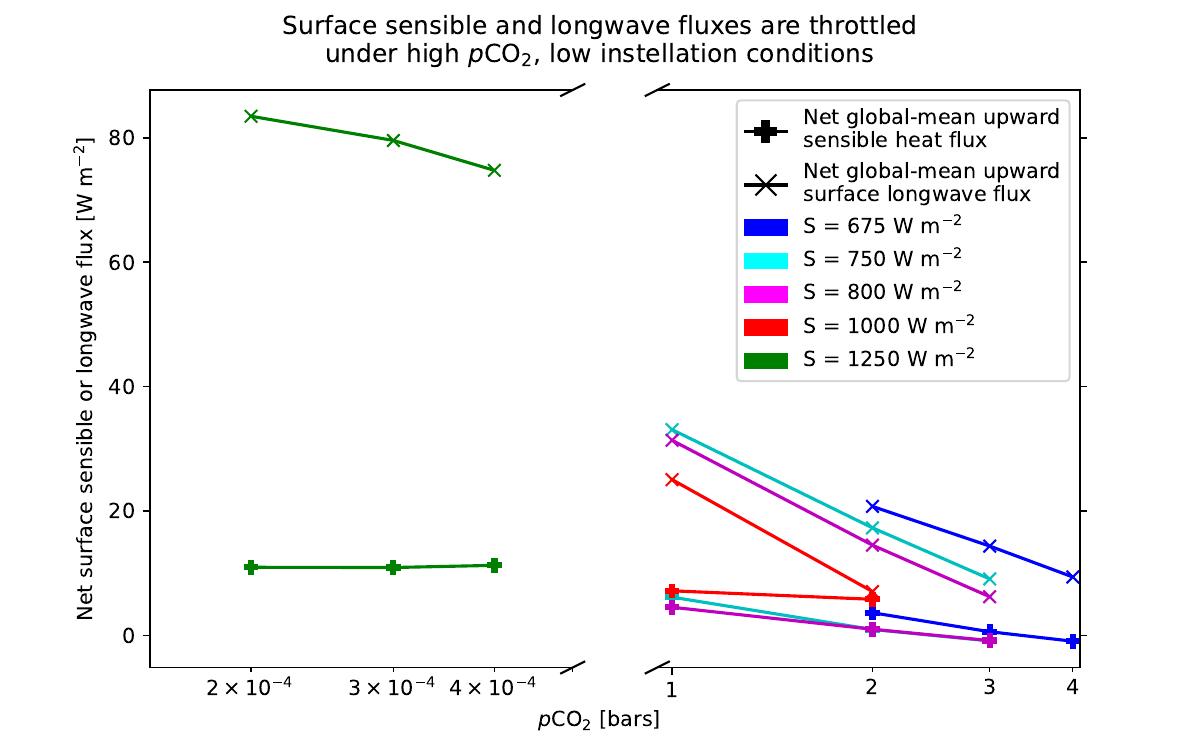}}
    \caption{\textbf{Net upward sensible  or longwave heat flux [W m$^{-2}$] as a function of $p$CO$_2$ for simulations at various $S$.} For low-$p$CO$_2$ simulations, the sensible heat flux (x-shaped markers) changes little with $p$CO$_2$, while for high-$p$CO$_2$ simulations it drops substantially, even becoming negative in some cases. For both low- and high-$p$CO$-2$ simulations, the longwave flux (plus-shaped markers) decreases with increasing CO$_2$, as expected. Colors are as in Fig. \ref{fig:ch4_ECS}.}
    \label{fig:ch4_longwave_sensible}
\end{figure*}

\begin{figure*}[htb!]
    \centering
    \makebox[\textwidth-50pt][c]{\includegraphics[width = \textwidth-50pt,keepaspectratio]{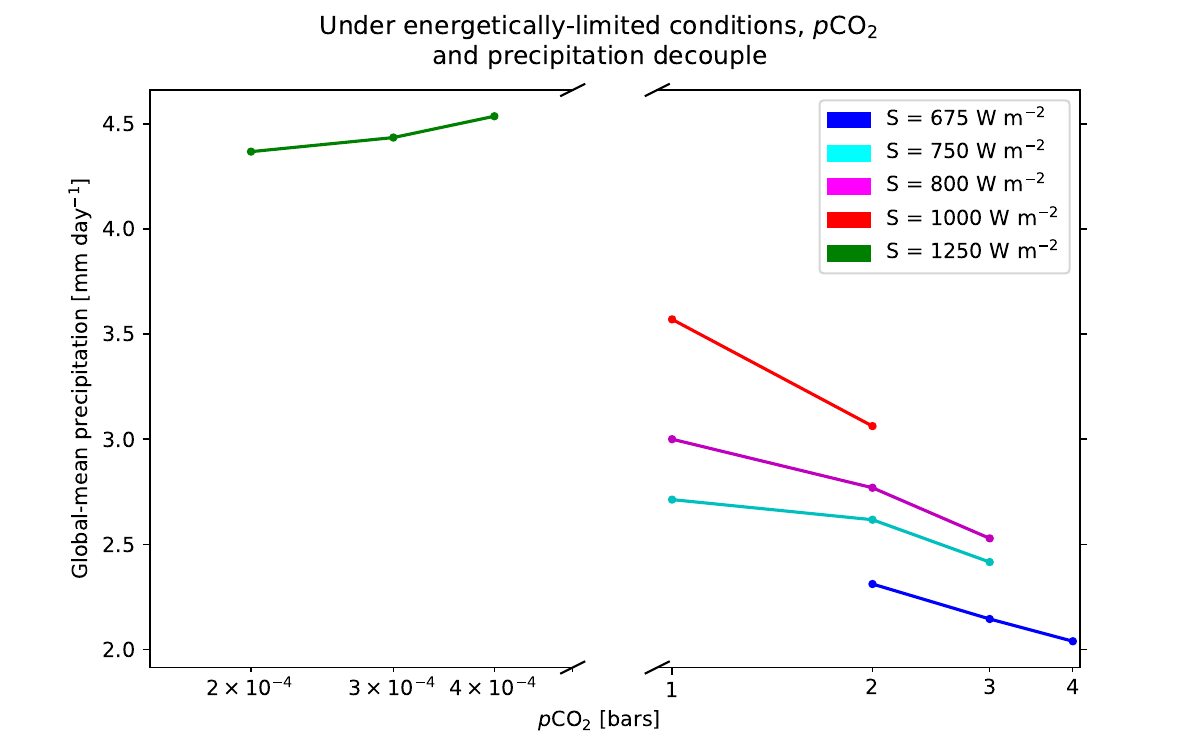}}
    \caption{\textbf{$p$CO$_2$ vs. global-mean precipitation.} For low-$p$CO$_2$, high-instellation simulations, precipitation increases with $p$CO$_2$. For high-$p$CO$_2$, low-instellation simulations in the regime where a large majority of their absorbed instellation goes into driving evaporation, precipitation falls with $p$CO$_2$. Colors are as in Fig. \ref{fig:ch4_ECS}.}
    \label{fig:ch4_pco2_precip}
\end{figure*}

\begin{figure*}[htb!]
    \centering
    \makebox[\textwidth-50pt][c]{\includegraphics[width = \textwidth-50pt,keepaspectratio]{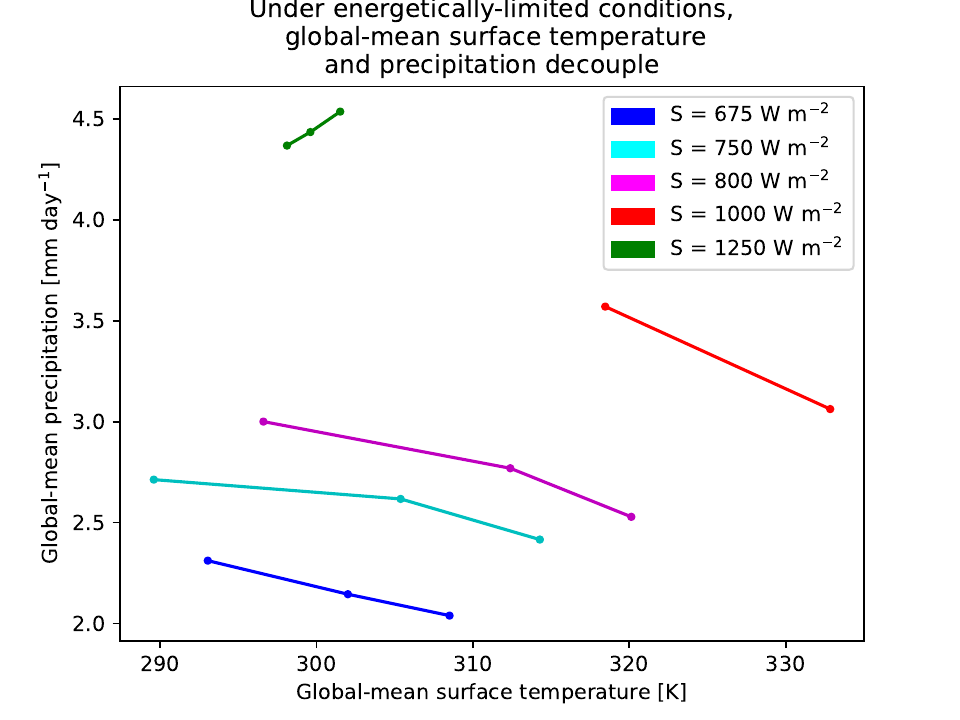}}
    \caption{\textbf{Global-mean surface temperature vs. global-mean precipitation.} For low-$p$CO$_2$, high-instellation simulations, precipitation increases with surface temperature. For high-$p$CO$_2$, low-instellation simulations in the regime where a large majority of their absorbed instellation goes into driving evaporation, precipitation falls with temperature. Colors are as in Fig. \ref{fig:ch4_ECS}.}
    \label{fig:ch4_temp_v_precip}
\end{figure*}

The behavior of the low-$p$CO$_2$ simulations is largely consistent with expectations from simulations of Earth climate. They approximately reproduce the expected positive trend in global-mean surface latent heat flux (green line with circle markers in Fig. \ref{fig:ch4_energetics}), the expected negative trend in global-mean upward longwave from the surface (green line with x markers in Fig. \ref{fig:ch4_longwave_sensible}), and the expected positive trend in precipitation (green lines in Figs. \ref{fig:ch4_pco2_precip} and \ref{fig:ch4_temp_v_precip}). Averaged across the three low-$p$CO$_2$ simulations, global-mean latent heat flux and global-mean precipitation both increased by 1.1 $\%$ K$^{-1}$, a bit less than the median value of 1.7$\%$ K$^{-1}$ found in Earth climate simulations \cite{held2006robust}. Despite increasing somewhat with surface temperature, for all three low-$p$CO$_2$ simulations the latent heat flux remained far below the limit set by $S_{\rm abs}$ (gre
en line with star markers in Fig. \ref{fig:ch4_energetics}), which decreased slightly in response to CO$_2$'s growth because of increased atmospheric absorption of instellation by H$_2$O as specific humidity rose. The total global-mean net upward longwave flux decreased by 8.7 W m$^{-2}$ from the 200 ppmv simulation to the 400 ppmv simulation, a somewhat greater reduction than the 4-5 W m$^{-2}$ per CO$_2$ doubling in Earth GCM simulations \citep[e.g.][]{gutowski1991surface,Boer93}, but this is to be expected since our low $p$CO$_2$ simulations have a warmer background climate than Earth, meaning they should experience larger increases in highly opaque lower-tropospheric water vapor for a given change in temperature than cooler Earth simulations because of the exponential dependence of H$_2$O partial pressure on temperature. The sensible heat flux of the low-$p$CO$_2$ simulations changes very little as a function of CO$_2$, with a reduction of $\approx$1 W m$^{-2}$ between the 200 pp
 mv and 300 ppmv simulations, and an increase of $\approx$1 W m$^{-2}$ between the 300 ppmv and 400 ppmv simulations (green line with plus-shaped markers in Fig. \ref{fig:ch4_longwave_sensible}). This contrasts slightly with simulations of Earth, where the surface's upward sensible heat flux often decreases by $\approx$1 W m$^{-2}$ under a doubling of CO$_2$ \citep{gutowski1991surface,Boer93,gomez2018climate}, consistent across models with fixed sea surface temperatures, slab oceans (like ours), and fully coupled dynamical oceans \citep{myhre2017pdrmip,myhre2018sensible}. This difference may be due to our neglect of topography, or it may be due to the fact that the climates we simulate are slightly warmer than modern Earth, biasing them toward smaller changes in sensible heat flux with surface temperature \citep{siler2019revisiting}. Nonetheless the changes in sensible heat flux are small in magnitude and the  flux itself is already a small enough term in the surface budget that it h
 as no impact on our qualitative results. 

\subsubsection{High $p$CO$_2$, low instellation}
The high $p$CO$_2$, low-instellation simulations display markedly different behavior from that described above and are clearly operating in a regime where the energetic limit set by absorbed instellation exerts substantial influence. For all sets of high-$p$CO$_2$ simulations irradiated by a given $S$, global-mean upward latent heat flux at the surface decreases with increasing $p$CO$_2$, despite the fact that surface temperature increases substantially (see blue, cyan, magenta, and red lines with circles in Fig. \ref{fig:ch4_energetics}). These reductions in latent heat flux are accompanied (and driven) by large reductions in absorbed instellation at the surface (see blue, cyan, magenta, and red lines with stars in Fig. \ref{fig:ch4_energetics}) due mostly to increases in albedo from CO$_2$'s Rayleigh scattering, from values of 0.16-0.17 for the $p$CO$_2$=1 bar cases, to 0.26-0.27 in the $p$CO$_2$=3 bars simulations, to 0.30 in the $p$CO$_2$=4 bars case, consistent with prev
ious simulations of planets with extremely high $p$CO$_2$ \citep{KASTING:1986p3082,ramirez2014can}. As $p$CO$_2$ is increased in these simulations, even though the absolute latent heat flux goes down, an increasing proportion of the absorbed instellation at the surface goes into evaporation, with values ranging from 66$\%$ to 92$\%$ for the high-$p$CO$_2$ cases, compared with a range of 57$\%$ to 60$\%$ for the low-$p$CO$_2$ simulations (see Table \ref{tab:ch4_energy_fractions}). Correspondingly, the sensible and longwave heat flux contributions to the surface energy budgets in the high-$p$CO$_2$ simulations are substantially reduced in both absolute and relative terms (see blue, cyan, magenta, and red plus-shaped markers and x-shaped markers in Fig. \ref{fig:ch4_longwave_sensible}). Interestingly, the global-mean sensible heat flux becomes negative (directed into the surface) in the $p$CO$_2$=3 bars simulations irradiated by $S=750$ W m$^{-2}$ and 800 W m$^{-2}$ and in the $p$CO$_2$
 =4 bars, $S=675$ W m$^{-2}$ simulation, but in each case the net sensible heating of the surface is less than 1 W m$^{-2}$.

As implied by the trends in latent heat flux vs $p$CO$_2$ displayed in Fig. \ref{fig:ch4_energetics}, global-mean precipitation rates in the high-$p$CO$_2$ simulations fall with increasing CO$_2$ (see blue, cyan, magenta, and red lines in Fig. \ref{fig:ch4_pco2_precip}). This results in counterintuitive behavior, driving precipitation rates that decrease with increasing surface temperature (see blue, cyan, magenta, and red lines in Fig. \ref{fig:ch4_temp_v_precip}). Despite some high-$p$CO$_2$ simulations having surface temperatures much larger than the low-$p$CO$_2$ simulations, the low-$p$CO$_2$ simulations all have global-mean precipitation rates larger than any of the high-$p$CO$_2$ simulations, consistent with the fact that the low-$p$CO$_2$ precipitation rates are sustained by latent heat fluxes larger than the $S_{\rm abs}$ received by any of the high-$p$CO$_2$ simulations except the $p$CO$_2$=1 bar, $S=1000$ W m$^{-2}$ case (compare green line with circles to blue, cy
an, magenta, and red lines with stars in Fig. \ref{fig:ch4_energetics}). In the next section we explore the implications of this energetically-limited precipitation behavior for the functioning of the carbon cycle.

\begin{table*}[htb!]
\centering
\begin{tabular}{rcccccc}
\hline
&\vline&\textbf{$\mathbf{\mathit{S}}$ [W m$^{-2}$]:}&&&\\
&\vline&675&750 & 800 & 1000 & 1250\\
\hline
\textbf{$\mathbf{\mathit{p}}$CO$_{2}$ [bars]:}&\vline&&&&\\
\textcolor{green}{2$\times10^{-4}$}&\vline&&&&&0.57\\
\textcolor{green}{3$\times10^{-4}$}&\vline&&&&&0.58\\
\textcolor{green}{4$\times10^{-4}$}&\vline&&&&&0.60\\
\cline{2-7}
1             &\vline  &&0.66&0.71&0.76&\\
2             &\vline &0.74&0.80&0.83&0.86&\\
3             &\vline &0.80&0.89&0.92&\\
4             &\vline &0.87&&&&\\
\hline
\end{tabular}
\caption{\textbf{Fraction of absorbed instellation going into evaporation ($\frac{L}{S_{\rm abs}}$)}. Table layout identical to Table \ref{tab:ch4_gcm} except $\frac{L}{S_{\rm abs}}$ values replace checkmarks.}
\label{tab:ch4_energy_fractions}
\end{table*}

\subsection{Weathering rates}\label{subsec:ch4_weathering_results}
In this Section we will present estimates of weathering rates based on output fields from the GCM simulations described above, with weathering fluxes calculated according to both the WHAK and MAC formulations. Sensitivity tests varying the parameters in the weathering models within plausible ranges generated fairly large changes in absolute fluxes but did not affect the qualitative picture presented here. 
\subsubsection{WHAK weathering}
\begin{figure*}[htb!]
    \centering
    \makebox[\textwidth-50pt][c]{\includegraphics[width = \textwidth-50pt,keepaspectratio]{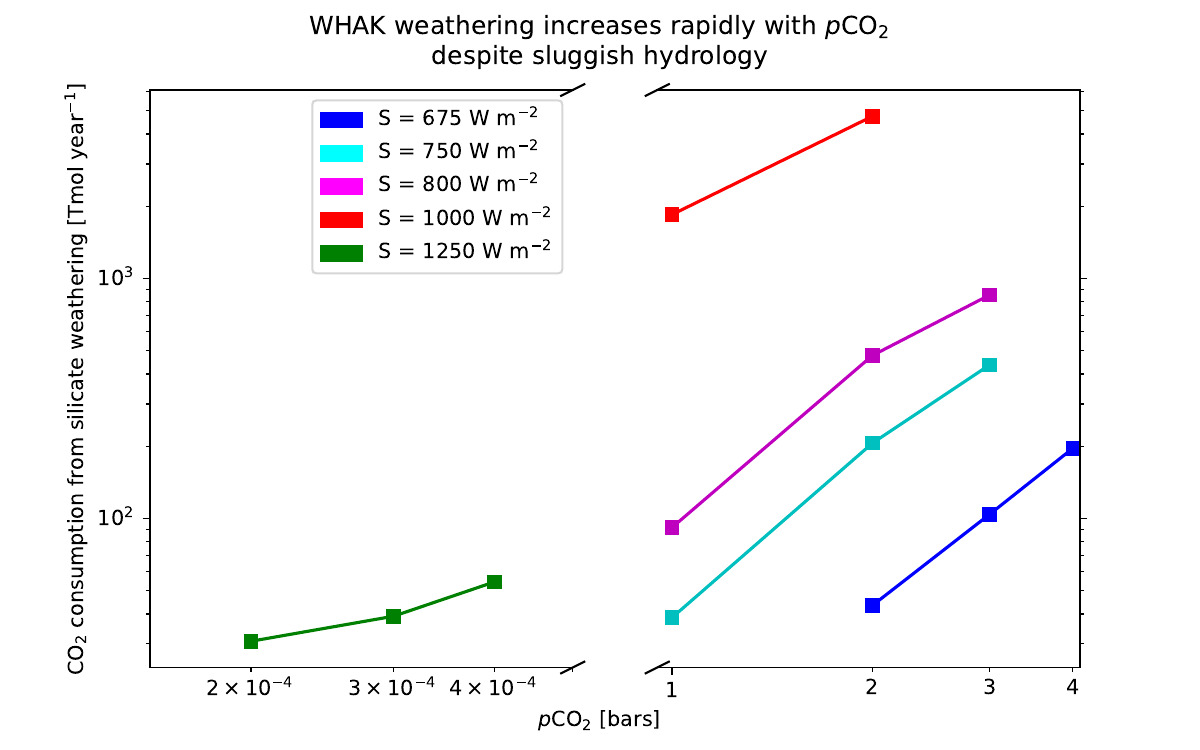}}
    \caption{\textbf{Global WHAK weathering rates [Tmol yr$^{-1}$] vs. $p$CO$_2$ for a variety of instellations}. Weathering fluxes are calculated according to the WHAK formulation (equation \ref{eqn:ch4_whak}). For all simulations, the weathering rate increases with CO$_2$. Colors are as in Fig. \ref{fig:ch4_ECS}.}
    \label{fig:ch4_whak}
\end{figure*}

When weathering is calculated according to the WHAK formulation (equation \ref{eqn:ch4_whak}), the weathering rate increases strongly in response to increases in $p$CO$_2$ and $S$ for all simulations, regardless of initial $p$CO$_2$ or instellation (see Fig. \ref{fig:ch4_whak}). This makes sense, given the WHAK formulation's exponential temperature-dependence and power-law $p$CO$_2$ dependence, both of which naturally produce big increases in weathering fluxes for modest changes in temperature and/or $p$CO$_2$. The increases produced by this model are unrealistically large even if we take the WHAK formulation at face value, as the global rate of rock uplift sets a ``supply limit'' on the rate of silicate weathering of $\mathcal{O}$(100) trillion moles of CO$_2$ per year on Earth [Tmol yr$^{-1}$] \citep[e.g.][]{kump2018prolonged}, but the positive slope of the curves in Fig. \ref{fig:ch4_whak} confirms in principle the continued operation of WHAK weathering as a negative feedb
ack even under high-$p$CO$_2$, low-instellation conditions where energetically-limited precipitation is relevant. Earth's outgassing rate is generally estimated to lie somewhere between 3 and 15 Tmol yr$^{-1}$ \citep{coogan2020average}, substantially smaller than even the smallest weathering rate produced by the these simulations, but this is due to our mostly arbitrary choice of weathering constant $W_{\rm ref}$ in equation \ref{eqn:ch4_whak}. A smaller $W_{\rm ref}$ would produce proportionally smaller weathering rates without changing the qualitative trends in weathering vs. $p$CO$_2$.

\subsubsection{MAC weathering}
\begin{figure*}[htb!]
    \centering
    \makebox[\textwidth-50pt][c]{\includegraphics[width = \textwidth-50pt,keepaspectratio]{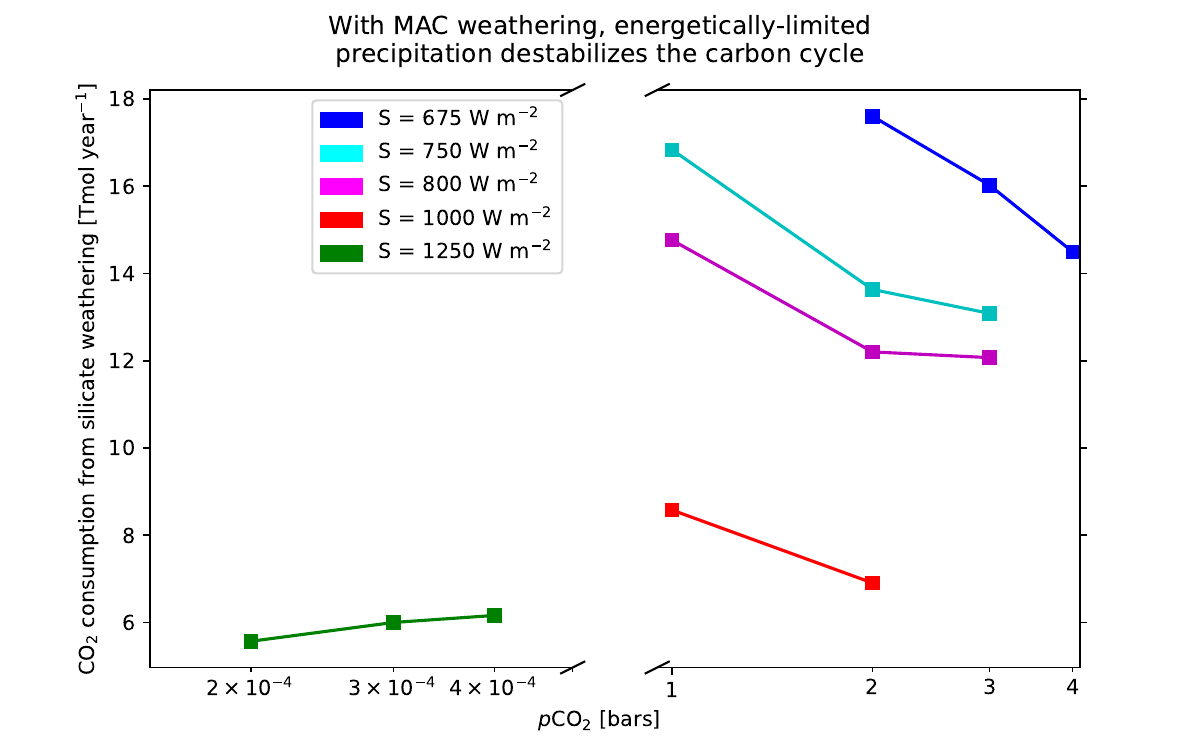}}
    \caption{\textbf{Global MAC weathering rates with fiducial parameter values [Tmol yr$^{-1}$] vs. $p$CO$_2$ for a variety of instellations}. Weathering fluxes are calculated according to the MAC formulation (equation \ref{eqn:ch4_mac}). For low-$p$CO$_2$ simulations, the weathering rate increases with CO$_2$, while for high-$p$CO$_2$ simulations weathering decreases with increasing CO$_2$. Colors are as in Fig. \ref{fig:ch4_ECS}.}
    \label{fig:ch4_mac}
\end{figure*}

With MAC weathering, the low $p$CO$_2$ simulations respond essentially as expected (Fig. \ref{fig:ch4_mac}). Although the strength of the weathering response to the changes in the surface climate is much weaker than in the WHAK case, the weathering rate in the low $p$CO$_2$ simulations still increases with increasing $p$CO$_2$, producing a stabilizing negative feedback. This provides tentative evidence that MAC-style continental weathering may be able to stabilize the climates of planets under Earth-like conditions of high instellation and low $p$CO$_2$.
\begin{figure*}[htb!]
    \centering
    \makebox[\textwidth-50pt][c]{\includegraphics[width = \textwidth-50pt,keepaspectratio]{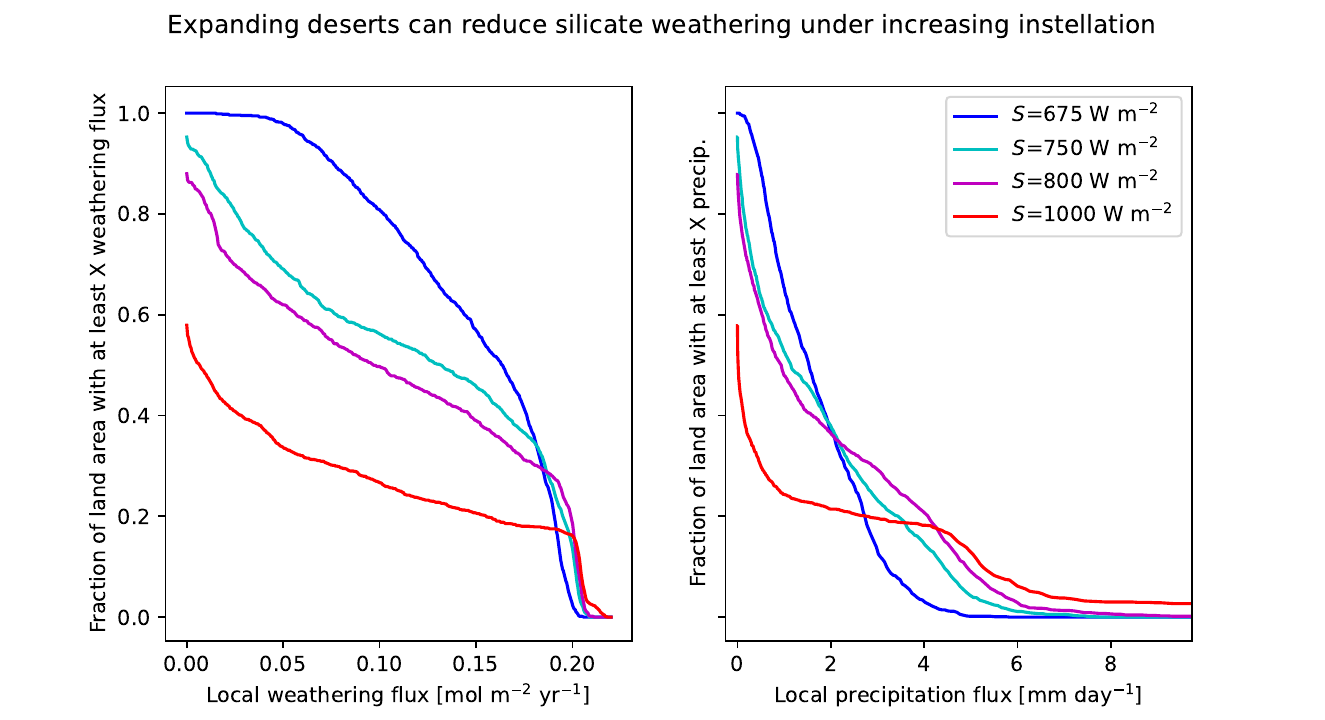}}
    \caption{\textbf{The change in spatial distribution of MAC weathering (left) and precipitation (right) that accompanies increasing TOA instellation} Results shown are for simulations with $p$CO$_2$ = 2 bars and $S=675$ W m$^{-2}$ (blue), $S=750$ W m$^{-2}$ (cyan), $S=800$ W m$^{-2}$ (magenta), and $S=1000$ W m$^{-2}$ (red). The y-value of a curve represents the fraction of land with at least as much weathering flux (left) or precipitation (right) as the corresponding x-axis value. Even though the total global precipitation increases as $S$ increases, the total global weathering rate decreases because less land is receiving substantial precipitation. See Section \ref{subsec:ch4_weathering_results}. Colors are as in Fig. \ref{fig:ch4_ECS}.}
    \label{fig:ch4_deserts}
\end{figure*}

In contrast, the high $p$CO$_2$, low instellation simulations in the energetically-limited regime produce a drastically different carbon cycle than the WHAK calculations or the low-$p$CO$_2$ MAC calculations would suggest. While they do display somewhat higher weathering rates than the low $p$CO$_2$ simulations due to the thermodynamic $p$CO$_2$-dependence of the maximum concentration of weathering products in runoff ($C_{\rm eq}$ in equation \ref{eqn:ch4_mac}, a much weaker effect than that of the $p$CO$_2$-dependence of WHAK weathering represented by $\beta$ in equation \ref{eqn:ch4_whak}), all of the high-$p$CO$_2$ simulations display the opposite response to further increases in CO$_2$: at a given $S$, as $p$CO$_2$ increases (along with surface temperature), the global weathering rate goes down (see blue, cyan, magenta, and red lines in Fig. \ref{fig:ch4_mac}). As described in Section \ref{subsec:ch4_weathering_models}, anti-correlation between global weathering rates and
 $p$CO$_2$ destabilizes the carbon cycle, so the high-$p$CO$_2$ simulations are all displaying a defective climate thermostat with a positive feedback that would likely induce runaway climate heating or cooling. Although this result runs counter to Earth-based intuitions, the unusual trends in $p$CO$_2$ vs. weathering make sense given the increased importance of hydrologic cycling in the MAC weathering framework and the reduction in global-mean precipitation triggered by reduced $S_{\rm abs}$.

Another unexpected weathering trend appears in the set of high $p$CO$_2$ simulations: holding CO$_2$ constant, the global weathering rate goes down even as instellation rises and drives both surface temperature (Fig. \ref{fig:ch4_ECS}) and global-mean precipitation (Fig. \ref{fig:ch4_pco2_precip}) upward with it. This trend obviously cannot be explained by a global-mean argument, since all of the variables that directly control weathering are, in bulk, either increasing (i.e. $q$ and surface temperature) or staying the same (i.e. $p$CO$_2$)  as $S$ is increased. Instead, the explanation lies in the intensification of moisture extremes expected in warming climates from simple thermodynamic considerations  \citep{held2006robust,allan2020advances}, often referred to by the phrase ``wet gets wetter, dry gets drier'' (WGWDGD) in discussions of Earth climate \citep{allan2020advances} though over the past several years it has become clear that this basic picture does not hold as wel
l over land in Earth simulations as it does over the ocean \citep{byrne2015response, feng2015global,allan2020advances}. Although WGWDGD does not perfectly describe the moisture response over land in Earth simulations under the (comparatively) small climate perturbations projected for this century, it quite accurately describes the behavior of the precipitation in our high-$p$CO$_2$ simulations as $S$ is increased. 

In other words, although the total global precipitation flux increases with $S$ in our high-$p$CO$_2$ simulations, it also becomes concentrated onto a smaller area, meaning deserts expand as $S$ goes up. For example, in the $S=675$ W m$^{-2}$, $p$CO$_2$=2 bars case, over half of the planet's land area receives more than 1 mm day$^{-1}$ precipitation, whereas only about a quarter of the land area in $S=1000$ W m$^{-2}$ receives that much water (see right side of Fig. \ref{fig:ch4_deserts}). This is compensated by an increasing fraction of landmass with high precipitation fluxes, i.e. only $\approx$5$\%$ of the landmass in the aforementioned $S=675$ W m$^{-2}$ simulation exceeds 4 mm day$^{-1}$ of rain, but about 20$\%$ of the landmass in the $1000$ W m$^{-2}$ simulation meets that criterion, and higher $S$ generally leads to a longer ``tail'' of landmass with large local precipitation fluxes. As can be seen by comparing the weathering flux curves on the left side of Fig. \ref{fig:ch4_deserts} with the precipitation flux curves on the right, local weathering tracks local precipitation quite directly, so it might seem that the long tail of high precipitation over land should more than make up for the expansion in dry areas. However, past a point determined by balance between the kinetics of silicate dissolution and the timescale and surface area of contact between water and dissolving silicates, extremely high precipitation has diminishing returns in terms of weathering rates \citep{maher2014hydrologic}. If water flushes out a weathering zone rapidly enough, the system becomes kinetically limited and any further increases to the water's flow rate will simply change the degree of dilution of weathering products rather than increasing their flux. We can see this effect in Fig. \ref{fig:ch4_deserts} where all of the local weathering fluxes drop off abruptly around 0.2 mol m$^{-2}$ yr even though a direct scaling of weathering flux with water flux would suggest
  a long tail of much larger local fluxes for the $S=1000$ W m$^{-2}$ simulation. Interestingly, although the precipitation rate and $S_{\rm abs}$ both increase as $S$ goes up, simulations at a given $p$CO$_2$ still move progressively closer to the energetic limit set by $S_{\rm abs}$ (see Table \ref{tab:ch4_energy_fractions}), which may play a role in constraining the areal extent of precipitation over land and forcing the growth of desert regions. The potential for reduction in weathering with increased $S$ is another destabilizing influence in the carbon cycle, since even with a functional negative feedback at a given instellation (i.e. positive slope in weathering vs. $p$CO$_2$), this effect would force $p$CO$_2$ to higher equilibria at higher instellations, likely significantly reducing a planet's habitable lifetime under increasing host star luminosity.

\section{Discussion}\label{sec:ch4_discussion}

\subsection{Catastrophic carbon cycle hysteresis}\label{subsection:CarbonHysteresis}
With MAC-style, hydrologically-regulated weathering, energetically-limited precipitation can cause a breakdown of the negative feedback on climate that emerges within the carbonate-silicate cycle. Our high-$p$CO$_2$, low-instellation simulations displayed reductions in weathering with CO$_2$ growth, opposite the trend required to stabilize climate by balancing CO$_2$ sequestration against CO$_2$ outgassing. This implies that in the portion of the instellation-$\mathrm{CO_2}$ space we have probed, which are generally characteristic of the outer reaches of the conventional HZ, coupled climate/carbon-cycle equilibria are unstable.  The instability is not just a local instability to infinitesimal displacements. The unstable equilibria represent the attractor basin boundary between low $\mathrm{CO_2}$ states when the system is displaced on the cold side, and states with very high $\mathrm{CO_2}$ when the system is displaced on the warm side.  The unstable branch extends over the e
ntire range of $\mathrm{CO_2}$ for which weathering decreases with $\mathrm{CO_2}$. For each instellation considered, weathering decreases with increasing $\mathrm{CO_2}$ in the high $\mathrm{CO_2}$ regime. However, in the colder low $\mathrm{CO_2}$ regime, weathering is expected to increase with $\mathrm{CO_2}$ because precipitation is not subject to an energy limit. (We have demonstrated that explicitly only for one instellation value). Based on these two end-member behaviors, it is inferred that there is a maximum weathering rate which, for the lower ranges of instellation we have considered, occurs at a $p\mathrm{CO_2}$ somewhere below 1 bar.  

\begin{figure*}[htb!]
    \centering
    \makebox[\textwidth][c]{\includegraphics[width = \textwidth,keepaspectratio]{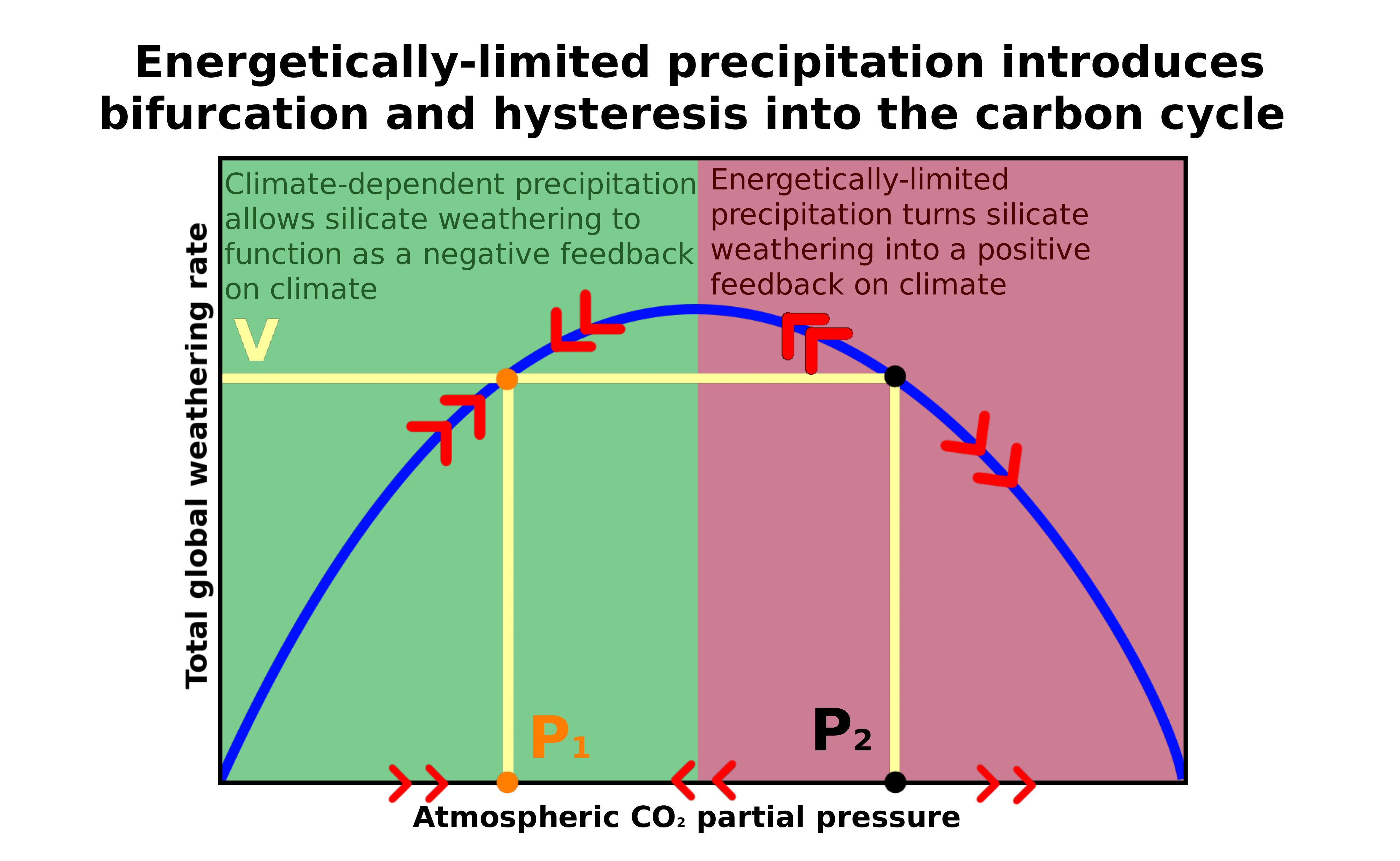}}
    \caption{\textbf{A schematic of the saddle node bifurcation in the carbon cycle suggested by our simulations.} The blue curve represents the global weathering rate as a function of $p$CO$_2$. Weathering increases with increasing $p$CO$_2$ (and surface temperature) in the green zone, and it decreases with increasing $p$CO$_2$ (and surface temperature) in the red zone. Weathering rates are equal to outgassing ($V$) at $p$CO$_2$=$P_1$ (orange) and at $p$CO$_2$=$P_2$ (black). The $P_1$ equilibrium is stable because weathering acts as a negative feedback in that regime, while the $P_2$ equilibrium is unstable due to silicate weathering's positive feedback behavior. The red arrows indicate the directions CO$_2$ and weathering would evolve for a climate intialized from various points relative to the two equilibria.}
    \label{fig:ch4_energetic_illustration}
\end{figure*}

The inferred structure of the weathering curve governing climate/carbon equilibrium is shown schematically in Fig. \ref{fig:ch4_energetic_illustration}, assuming there to be only one local maximum of weathering.  The value of $p\mathrm{CO_2}$ at which peak weathering occurs, and the height of the peak, depend on instellation. Our simulations have probed just a portion of the relevant instellation space, and only a portion of the right-hand half of the curve where weathering decreases with $p\mathrm{CO_2}$. In particular, we have not located the peak of the weathering curve, which occurs in the gap between our low and high $p\mathrm{CO_2}$ simulations. The geometry of the weathering curve indicates the presence of a saddle-node bifurcation  \citep[e.g.][]{Strogatz-1994:nonlinear} in the climate/carbon equilibrium. For a given volcanic outgassing rate below peak weathering (indicated by the horizontal line labeled "V" in the figure), there are two equilibria. The equilibrium $P
_2$ to the right is unstable; this is the unstable high $p\mathrm{CO_2}$ equilibrium we have identified in our simulations. If the system initially has $p\mathrm{CO_2}$ anywhere to the left of $P_2$, it will be attracted to the stable low $p\mathrm{CO_2}$ equilibrium $P_1$ Because our simulations have not located the peak, we cannot determine whether the stable state $P_1$ is a Snowball or a habitable low $p\mathrm{CO_2}$ state with above-freezing conditions. 

If the system starts somewhat to the right of $P_2$, then $\mathrm{CO_2}$ will accumulate until something arrests the process at higher $p\mathrm{CO_2}$. This could terminate at complete degassing of the interior $\mathrm{CO_2}$ reservoir, leading to a hot Venus-like state with a thick gaseous atmosphere, or a more temperate state where $\mathrm{CO_2}$  accumulates in the form of a liquid ocean, depending on instellation as in \citet{graham2022co2}. Alternately, it is possible that the weathering curve turns around at sufficiently high $p\mathrm{CO_2}$, allowing a new stabke high $p\mathrm{CO_2}$ equilibrium to exist, though possibly at an uninhabitably hot temperature.  Because the processes leading to the decrease of weathering with $p\mathrm{CO_2}$ persist and even accentuate at higher values than we have probed, we think this unlikely, but it is a possibility that cannot be ruled out at present. In any event, the unstable equilibrium $P_2$ is a basin boundary between stab
le low and high $p\mathrm{CO_2}$ states. 

It is generic to the geometry of saddle-node bifurcations that the system supports hysteresis. Suppose the planet starts in the stable equilibrium $P_1$ at a time when volcanic outgassing is relatively weak. If outgassing increases subsequently, and exceeds the peak of the weathering curve, then the system will transition to a stable high $p\mathrm{CO_2}$ equilibrium somewhere past the right edge of the diagram.  Because any state to the right of $P_2$ will be attracted to the high $p\mathrm{CO_2}$ state, volcanic outgassing would need to be reduced to below its original value in order to restore a stable low $p\mathrm{CO_2}$ state. Because the peak of the weathering curve depends on instellation, it is also possible for such hysteresis loops to occur for fixed outgassing, as a result of evolution of stellar luminosity. Our simulations indicate that increasing instellation in the high $p\mathrm{CO_2}$ states decreases weathering, and so likely reduces the peak weathering. Thi
s scenario could lead to a new form of habitability termination as a G or F star increases in luminosity over its main sequence lifetime. 

There are additional pathways into the energetically-limited regime that do not require any changes to outgassing or instellation. For example, if the $p$CO$_2$ necessary to deglaciate a planet in a snowball state lies at $p$CO$_2>P_2$ (black markers in Fig. \ref{fig:ch4_energetic_illustration}), then escape from global glaciation due to reduced weathering in icy conditions could counter-intuitively propel a previously temperate planet into a death spiral of uncontrolled heating. Upon accumulating enough CO$_2$ to deglaciate, rather than weathering away the excess to return to a temperate climate, runaway CO$_2$ accumulation would ensue due to the processes described above for planets initialized from $p$CO$_2>P_2$. Similarly, if Earth-like planets outgas massive CO$_2$ atmospheres during their magma ocean phases like some models suggest \citep[e.g.][]{solomatova2021genesis}, some may become ``locked in'' to this early, hot, high-$p$CO$_2$ phase, since their weathering rates 
after magma ocean crystallization and water ocean condensation might remain pinned below their early outgassing rates due to energetically-constrained precipitation. However, expectations that planets will also face a large flux of impacts producing highly weatherable ejecta in the early stages after planet formation \citep[e.g.][]{kadoya2020probable} may help mitigate this danger. Further, at low instellations, deglaciation can be prevented by CO$_2$ condensation onto the icy planetary surface \citep{turbet2017co,kadoya2019outer}, which may restrict the parameter space that allows for a post-Snowball death spiral. 

A more quantitative example of how CO$_2$ runaway might play out can be illustrated with direct reference to Fig. \ref{fig:ch4_mac}. The $p$CO$_2$=1 bar, $S=1000$ W m$^{-2}$ simulation (red squares in Fig. \ref{fig:ch4_mac}) displays a weathering rate of 8.6 Tmol yr$^{-1}$. If we take this as our initial climate state and impose a volcanic CO$_2$ degassing rate of 10 Tmol yr $^{-1}$, which is well within the range of estimates for Earth's outgassing rate \citep{catling2017atmospheric,coogan2020average}, the planet's $p$CO$_2$ will begin to grow because the flux into the atmosphere is greater than the flux out. Under Earth-like conditions, this would cause warming, an intensification of the hydrological cycle, and an acceleration in weathering (as we see in the low-$p$CO$_2$ cases, the green squares in Fig. \ref{fig:ch4_mac}), moving the carbon cycle closer to equilibrium, but in this case the opposite happens, with a progressive reduction in the weathering rate as CO$_2$ accu
mulates. By the time our hypothetical $S=1000$ W m$^{-2}$ planet had accumulated another bar of CO$_2$, its weathering rate would have fallen to only 6.9 Tmol yr$^{-1}$ (see red square at $p$CO$_2$=2 bars in Fig. \ref{fig:ch4_mac}). Thus the planet would have a carbon cycle further out of balance with its 10 Tmol yr$^{-1}$ outgassing rate than when it started, with no signs of slowing down. We did not carry the $S=1000$ W m$^{-2}$ simulations to higher $p$CO$_2$ because of the temperature constraints of our modeling framework, but even at $p$CO$_2$=2 bars the planet is verging on inhospitable conditions, with a global-mean surface temperature in excess of 330 K (red dot at $p$CO$_2$=2 bars in Fig. \ref{fig:ch4_ECS}). Assuming the negative weathering trend holds out to even larger $p$CO$_2$, or at least that the weathering rate remains below the assumed 10 Tmol yr$^{-1}$ outgassing rate, the planet would eventually be forced by the carbonate-silicate cycle into an uninhabitably hot st
 ate, likely continuing to warm until it degassed all of the available CO$_2$ from its interior, ending up with a dense, steamy, supercritical CO$_2$/H$_2$O atmosphere unless some unknown process was triggered along the way that allowed the weathering rate to accelerate back to parity with the outgassing rate. Simulations of water photolysis and hydrogen escape on CO$_2$-rich planets orbiting G-stars suggest that the upper-atmospheric cold trap generated by the low instellations and high CO$_2$ partial pressures under consideration would throttle water loss to insignicant levels, allowing these hot, steam-rich surface conditions to persist over geologic timescales \citep{wordsworth2013water}.

At somewhat lower instellations than $S=1000$ W m$^{-2}$, CO$_2$ condensation at the surface would become a possibility at high enough $p$CO$_2$, meaning the energetic limit on precipitation presents a pathway to reach the stable CO$_2$ ocean states proposed in \citet{graham2022co2}. The mechanism of carbon cycle destabilization suggested in that study was based on the fact that CO$_2$ becomes a coolant at extremely high partial pressures, which also leads to a slowdown of weathering with $p$CO$_2$ accumulation but requires much higher CO$_2$ to initiate than the mechanism discussed here. This means that energetically-limited weathering may increase the probability that low-instellation planets within the HZ end up with CO$_2$ oceans, with uncertain implications for their habitability. Although CO$_2$ ocean worlds are restricted to relatively clement climates below 304 K (the critical temperature of CO$_2$), they would still sport extreme surface pressures (up to 72 bars), an
d in the process of accumulating that much CO$_2$ surface temperatures can peak at extreme levels ($>400$ K) before Rayleigh scattering begins to outweigh CO$_2$'s greenhouse effect and further accumulation begins to cool the planet \citep{graham2022co2}. Intriguingly, some work has suggested that liquid and supercritical CO$_2$ may be conducive to prebiotic chemistry \citep{shibuya2022liquid}, and supercritical CO$_2$ may drive rapid carbonate formation \citep[see Section \ref{subsec:ch4_escape_routes} and][]{mcgrail2017field}, suggesting these worlds may not be as hostile to life as they appear at first.

s debate remains over proxy interpretation \citep{galili2019geologic,verard2019plate,herwartz2021co2}, some paleotemperature constraints suggest that the Archean Earth may have had surface temperatures in the vicinity of 340 K \citep[e.g.][]{lowe2020constraints,mcgunnigle2022triple}, again presenting the possibility that early Earth may at times have verged on the energetically-limited weathering behavior simulated here. Finally, we note that early Mars, which was irradiated by a lower instellation than any of the planets we modeled, is generally postulated to have had a CO$_2$-dominated atmosphere, and displays abundant evidence of active hydrologic cycling \citep[e.g.][]{kite2019geologic,kite2019persistence,kite2021warm}, may be another environment in which energetically-limited precipitation and weathering became relevant. 

\subsection{Limit cycling?}
First, we note that our WHAK weathering results (Fig. \ref{fig:ch4_whak}) are qualitatively consistent with previous WHAK-based exoplanet weathering calculations driven by zero-dimensional \citep{Menou2015,abbot2016analytical}, one-dimensional \citep{haqq2016limit,kadoya2019outer}, and three-dimensional \citep{paradise2017} climate models. In all of these cases, the direct power-law dependence of weathering fluxes on $p$CO$_2$ (the $\beta$ term in eqn. \ref{eqn:ch4_whak}) means that planets at lower instellations (where larger $p$CO$_2$ is necessary to maintain a given surface temperature) display much higher weathering rates under temperate conditions, necessitating colder surface climates to reduce weathering enough to equilibrate with an Earth-like outgassing rate. Under fairly broad combinations of instellation and weathering properties \citep[e.g.][]{abbot2016analytical}, this direct $p$CO$_2$-dependence was suggested to draw CO$_2$ down to levels that would glaciate pla
nets. This led to the prediction that terrestrial planets in the outer reaches of the HZ would spend most of their time in snowball states, punctuated by intermittent periods of temperate surface conditions when cold- and ice-induced slowdowns in weathering allowed CO$_2$ to build up to high enough levels to achieve deglaciation briefly, at which point weathering would accelerate again and re-glaciate the planet, a periodic process termed ``limit cycling'' \citep{Menou2015,haqq2016limit}. The huge increases in WHAK weathering rates for our high-$p$CO$_2$ simulations (Fig. \ref{fig:ch4_whak}) are consistent with this picture, though we did not carry out simulations at low enough $p$CO$_2$ values and surface temperatures to determine whether the WHAK weathering rates would remain extremely high even as the planets approached glaciation. 

Compared to the results discussed above, our MAC weathering calculations (see Fig. \ref{fig:ch4_mac}) suggest completely different behavior for planets with CO$_2$ dominated atmospheres at reduced instellations within the HZ. Although they do display somewhat higher weathering rates than their low-$p$CO$_2$ counterparts because of the $p$CO$_2$-dependence of the equilibrium concentration of solutes ($C_{\rm eq}$ in equation \ref{eqn:ch4_mac}), the high-$p$CO$_2$ MAC simulations do not display the orders-of-magnitude increase in weathering compared to cases with lower CO$_2$ that is seen with WHAK. From this it seems that MAC weathering imparts much less susceptibility to the limit cycling mechanism, which at first glance appears to be a point in favor of climate stability at low instellation. However, the carbon cycle instability discussed in Section \ref{subsection:CarbonHysteresis}, which sets in under low instellation and high $\mathrm{CO_2}$ conditions when using MAC weat
hering, has the potential to pose an equal or greater threat to habitability in the outer portions of the conventional HZ. Unlike climate limit cycling, this instability does not depend on glaciation, though the cold-side attractor could in some cases be a Snowball state.  In such cases, the $\mathrm{CO_2}$ build-up required to trigger deglaciation could be large enough to flip the system into the hot-side high $\mathrm{CO_2}$ attractor.  Since we have not quantified the hot-side attractor, we have no basis at present to speculate as to whether there are mechanisms that could return the system to a Snowball state and thus lead to a form of climate limit cycling. 

The novel carbon cycle instabiity our work has identified suggests mechanisms whereby the geochemical HZ could be significantly contracted relative to the convential HZ which doesn't take into account geochemical constraints.  The extent to which habitability in the outer portions of the conventional HZ is actually threatened is subject, however, to the resolution of a number of caveats.

\section{Caveats} \label{subsec:ch4_escape_routes}
In this study, we focused exclusively on modeling continental silicate weathering rates in a significantly simplified cloud-free GCM with a specific, idealized land configuration and an assumption of spatially uniform lithology and soil properties. There are a variety of possibilities that we did not model which could prevent or complicate the carbon cycle destabilization discussed above. 

\subsection{Mind the gap}
The most important limitation in the results presented above is the gap between our low $p\mathrm{CO_2}$ simulations and our high $p\mathrm={CO_2}$ simulations. This gap was due to technical issues with finding a computationally feasible spectral representation that would enable the radiation code to cover the whole range. Filling in the gap also would require incorporation of ice-albedo feedback, and probably also dynamic ocean effects, in order to resolve the Snowball transition properly. 

For our lowest instellation case, 675$\mathrm{W/m^2}$, 1D calculations indicate that $p\mathrm{CO_2}$ in excess of 1 bar would be necessary to keep the global mean surface temperature above freezing \citep[Fig. 6][]{kopparapu2013revised}. Our simulations did not probe below 2 bars for this instellation, but at 2 bars the weathering still is strongly increasing as $p\mathrm{CO_2}$ decreases, so it is likely that the unstable feedback continues to 1 bar and below.  For this case at least, it seems plausible that the low $p\mathrm{CO_2}$ attractor is a Snowball.  For the higher instellations in our simulations, the actual position of the peak weathering becomes crucial to the question of whether the attractor is a Snowball. 

Our suite of simulations also has a gap for $p\mathrm{CO_2}$ above 4 bars, so we cannot say where the $p\mathrm{CO_2}$ and temperature ultimately land in circumstances where runaway accumulation occurs.  

An additional gap in our story is that we have not probed instellations below 675 $W/m^2$, whereas the outer edge of the conventional HZ for G stars is in the vicinity of 475$W/m^2$, requiring nearly 10 bars of $\mathrm{CO_2}$ for its maintainence. Aside from the instellation being lower than we have probed, the $\mathrm{CO_2}$ level is somewhat over twice the maximum value we considered.   We can not at present rule out the possibility that the climate/carbon equilibrium stabilizes as the outer HZ edge is approached. This in itself would be an interesting state of affairs, leading to the notion of "habitable bands" rather than a continuous HZ. 

At lower instellations than we have probed, which correspond to conditions nearer to the outer edge of the conventional
HZ, $\mathrm{CO_2}$ condensation will occur for high $\mathrm{CO_2}$ states, first in the upper atmosphere and then approaching the ground as instellation decreases.  This condensation has the dual effect on the planetary energy budget of keeping the upper atmosphere warmer than it would have been on the noncondensing adiabat, and  through the radiative effects of $\mathrm{CO_2}$ ice clouds.  $\mathrm{CO_2}$ ice clouds aloft have a cooling effect through increasing albedo, and a warming effect through their (infrared scattering) greenhouse effect. Regardless of whether the net effect is warming, cooling or neutral, the cloud albedo further reduces the surface instellation, making the energy limit more stringent than in clear sky conditions and thus further reducing weathering as $\mathrm{CO_2}$ increases. This would add to the destabilization of the climate-carbon equilibrium. 

\subsection{Seafloor weathering}
One important consideration is the potential contribution of low-temperature off-axis hydrothermal basalt alteration, frequently referred to as ``seafloor weathering,'' which has been proposed as an alternative or complementary stabilizing, temperature- and $p$CO$_2$-sensitive CO$_2$ sequestration flux analogous to the continental weathering feedback \citep[e.g.][]{francois1992modelling,Brady:1997p3530,coogan2013evidence,coogan2015alteration,krissansen2017constraining,krissansen2018constraining,coogan2018temperature,hayworth2020waterworlds}. Some work has discounted its ability to act as a stabilizing feedback \citep{CALDEIRA:1995p2285,abbot12-weathering}, but these conclusions have been questioned due to laboratory \citep{Brady:1997p3530} and geochemical \citep{coogan2013evidence,coogan2015alteration,coogan2018temperature} data that indicate an appreciable temperature-dependence of seafloor weathering reactions, potentially allowing the process to accelerate under warmer cli
mates and operate as a negative feedback, with important implications for the history of Earth \citep{krissansen2017constraining,krissansen2018constraining} and the habitability of exoplanets \citep{hayworth2020waterworlds,chambers2020effect}. However, these modeling works generally ignore the throttling effect of clay formation which plays such a fundamental role in the weakening of the continental silicate weathering feedback in the MAC framework. If the waters flowing through seafloor basalts tend to reach their equilibrium concentration of weathering products before the formation of carbonates, then without a feedback between porewater flow rates and global surface temperatures there is limited scope for seafloor weathering to operate as a thermostat. Developing a complete MAC-style model of seafloor weathering that accounts for the major controls on porewater flow rates and the precipitation of relevant clay phases, as suggested in \citet{graham2020thermodynamic}, with first ste
 ps taken in \citet{hakim2021lithologic}, is a necessary next step toward evaluating the importance of seafloor weathering to the carbon cycle of Earth and other exoplanets. This is particularly crucial given the fact that clay formation on the seafloor was likely much more efficient in early Earth's oceans (and in the oceans of abiotic exoplanets) due to the lack of biosilicifying organisms, which maintain the modern Earth's ocean in a subsaturated state with respect to silica \citep{siever1992silica,kalderon2021lithium}. Silica is one of the weathering products that drives the formation of many of the clays that consume weathering-derived cations and reduce carbonate precipitation, decreasing the effectiveness of weathering at driving carbon sequestration; thus, although seafloor clay precipitation (``reverse weathering'') has been suggested to exert its own form of negative feedback \citep[e.g.][]{isson2018reverse,krissansen2020coupled}, it may also prevent traditional seafloor we
 athering from operating efficiently. The net impact on the carbon cycle of weathering-related processes occurring on the seafloor remains unclear \citep{krissansen2020coupled}.

\subsection{Water clouds}
In future work, it will be important to determine whether water clouds could stabilize the MAC weathering feedback.  Water clouds affect the hydrology because their albedo further reduces the surface instellation. This is true for both boundary layer clouds and clouds aloft, but clouds aloft in addition exert a warming influence through their greenhouse effect, which can partly compensate or even overwhelm the cooling effect of the albedo.  We have cited some reasons to suspect that boundary layer water clouds may be absent in a high $p\mathrm{CO_2}$ regime, but if they are still present at, e.g., the 1 bar climate but dissipate as $p\mathrm{CO_2}$ increases further, than that would partly offset the energy limit effects on precipitation due to the albedo of $\mathrm{CO_2}$.  Insofar as MAC weathering is more sensitive to hydrology than to direct temperature effects, dissipation of high clouds could have the same effect on weathering, despite their warming influence.  In case
s where high clouds exert a dominant warming effect, though, the additional warming could enhance the desertification effect which in our simulations contributes to the destabilizing weathering feedback. Generally speaking, it should be noted that the incremental albedo effect of clouds is muted in climates where albedo is already high due to $\mathrm{CO_2}$, and that the greenhouse effect of clouds can be muted if they lie below the radiating level of an optically thick $\mathrm{CO_2}$ atmosphere.

\subsection{Continental configuration}
The amount and spatial arrangement of land can have complex and difficult-to-predict impacts on continental weathering rates via changes to runoff and precipitation \citep{baum2022sensitive}. It may be that certain continental configurations tend to produce positive feedback behavior, while others produce negative feedback behavior. Relatedly, topography heavily influences (and is influenced by) precipitation \citep{roe2005orographic} and erosion \citep{montgomery2002topographic} rates, which in turn affect the thickness and age of soils \citep[e.g.][]{heimsath2000soil,owen2011sensitivity}, with direct impacts on silicate weathering rates \citep{ferrier2008effects,hilley2010competition,west2012thickness,maher2014hydrologic}. Surface evolution driven by plate movements \citep[e.g.][]{coy2022diskworld} means that all of these factors may constantly co-evolve on an Earth-like planet with plate tectonics. Further, planets in the ``stagnant lid'' mode \citep[no plate tectonics, e.
g.][]{foley2019habitability} may display systematic differences in topograpy \citep{guimond2022blue} and outgassing rates \citep{guimond2021low}, likely introducing further major variation into all of these factors controlling weathering rates and climate evolution. Correlations and feedbacks between the many variables controlling weathering fluxes could significantly change the picture we have presented, and a much broader parameter sweep and investigation of these issues is necessary. 
\subsection{Ocean heat transport}\label{subsec:oht}
Modeling climate with a slab (as opposed to dynamical) ocean is a significant simplification, but we consider it unlikely to have a large impact on our qualitative results. Note first that our simulations are carried out without a seasonal cycle, so that the ocean response time is immaterial except insofar as it somewhat dampens surface temperature response to synoptic variability.  Additionally, study of Earth's climate suggests that oceans carry a relatively small proportion of meridional heat transport, with the bulk carried by the atmosphere \citep{trenberth2001estimates}. Moreover, there are robust reasons to expect the atmosphere to compensate for absence of ocean heat transport \citep{farneti2013meridional}. 

Nonetheless, ocean heat transport can increase atmospheric water vapor by spreading atmospheric convection out of a planet's deep tropics, resulting in global-mean warming \citep{herweijer2005ocean}. Furthermore, in an icy climate, ocean heat transports can have considerable warming effects because a small amount of heat delivered to and under the sea ice margin is very efficient at melting ice \citep[e.g.][]{rose2015stable}. The latter is not a factor in the climates we explore, since they are all too warm to support much ice, but it is well established that ocean heat transport and sea ice dynamics have a strong influence on the the $\mathrm{CO_2}$ concentration at which a planet transitions into a Snowball state \citep{Pierrehumbert-et-al-2010:neoprot}, generally requiring lower $\mathrm{CO_2}$ for global glaciation than is the case for slab models.  Inclusion of dynamical ocean and sea ice effects would be crucial in order to definitively answer whether the low $\mathrm{C
O_2}$ states the unstable system is attracted to on the cold side of the unstable equilibrium are Snowballs. This is a question we do not attempt to resolve in the present simulations. We also note that some simulations of exoplanetary habitability in the middle- and outer-reaches of the HZ have found ocean heat transport to have a major effect on climate by warming the nightsides of synchronously rotating exoplanets, where they do not receive instellation from their parent stars \citep{yang2013,yang2014,hu2014role,yang2019ocean}. However, since our study is concerned with rapidly rotating exoplanets, this mechanism is less crucial.
\subsection{Continental lithology}
We have considered only one mineral composition for the weatherable surface, but weathering behavior is sensitive to lithology \citep{hakim2021lithologic}. Additionally, the continental crust would typically exhibit considerable spatial variations in lithology, driven by a multitude of tectonic processes.  Such variations can accentuate the effect of continental configuration. 

A possible exit route from uncontrolled CO$_2$ accumulation could arise through the power-law dependence of $C_{\rm eq}$ in equation \ref{eqn:ch4_mac} (characterized by exponent $n$ in Table \ref{tab:ch4_weathering_values}), which allows for larger concentrations of weathering products in a given amount of water for planets with larger $p$CO$_2$, suggesting that further accumulation of CO$_2$ in the energetically-limited regime could eventually reverse the sign of the weathering curve back into negative feedback territory. This is the mechanism that allowed the subset of simulations that entered the parameterized energetically-limited precipitation regime in the global-mean study of \citet{graham2020thermodynamic} to avoid runaway CO$_2$ accumulation, instead equilibrating at hot temperatures with high $p$CO$_2$. In this scenario, the slope of the blue weathering curve in Fig. \ref{fig:ch4_energetic_illustration} would eventually reverse and begin to increase again at high $p
$CO$_2$, introducing a new, stable carbon cycle equilibrium at a $p$CO$_2>P_2$. However, we note that the chemical equilibrium constants governing $C_{\rm eq}$ are negatively temperature-dependent for many lithologies, such that the maximum concentration of weathering products decreases exponentially with increasing temperature, opposite to the behavior of kinetic rate constants and offsetting the power-law increase of $C_{\rm eq}$ with $p$CO$_2$ \citep{hakim2021lithologic}. A set of weathering calculations including this effect led to even larger drops in global weathering rates with $p$CO$_2$ in the high-CO$_2$ simulations, exacerbating the bifurcation and hysteresis identified above. Still, there remains a possibility that weathering rates could, under some circumstances, begin to climb again at extremely high $p$CO$_2$ when CO$_2$ begins to cool the planetary surface instead of warm it. Further, some field measurements suggest that supercritical CO$_2$ forms carbonates extremely 
 rapidly upon being brought into contact with basalts \citep{mcgrail2017field}, suggesting the intriguing possibility that weathering would increase greatly under supercritical conditions, providing another possible mechanism for the weathering curve to regain its negative feedback at extremely high $p$CO$_2$.

\subsection{Spin state}
This study focused on rapidly-rotating planets orbiting Sun-like G stars. Carbon cycling on slowly-rotating and tidally-locked planets has received much less attention. \citet{Kite:2011} found that WHAK-style silicate weathering can transform into a positive feedback for tidally-locked planets with thin, CO$_2$-dominated atmospheres through a mechanism completely different from that explored here. Weathering-induced climate hysteresis on tidally-locked planets with very limited water inventories has also been suggested \citep{ding2020stabilization}. Other WHAK-based calculations have suggested a significant dependence of weathering rates on planetary rotation rate \citep{jansen2019climates} and enormous changes to weathering rates as a function of continental position on fully synchronous rotators \citep{edson2012carbonate}. These effects are especially important since slowly-rotating planets are expected to experience very efficient true polar wander, leading to continuous r
eorientation of their surfaces with respect to the substellar point on timescales comparable to that of the carbonate-silicate cycle as a result of mantle convection \citep{leconte2018continuous}. Hydrological cycling on tidally-locked planets displays subtle behavior with unclear implications for weathering rates \citep{labonte2020sensitivity}. MAC weathering calculations have not yet been applied to tidally-locked climates, so it is unclear whether the mechanisms we have identified in this study will come into play in that context, but this is a crucial area for future research to evaluate the potential climate stability of planets orbiting M-dwarfs, the most plentiful stars in the universe \citep[e.g.][]{catling2017atmospheric}.

\section{Conclusion}\label{sec:ch4_conclusion}
In this study, we have calculated estimates of continental silicate weathering fluxes for Earth-like exoplanets by applying the MAC weathering model to output from GCM simulations of planetary climate under a variety of $p$CO$_2$ values and TOA instellations. Weathering rates and fluxes predicted according to MAC diverge profoundly from values calculated according to the more widely used WHAK model, particularly at lower instellations within the HZ. We have shown that for a considerable range of low instellations and high $p\mathrm{CO_2}$ generally characteristic of the outer portions of the conventionally defined habitable zone, the common assumption that silicate weathering provides a stabilizing feedback on climate can break down, because the climate/carbon-cycle equilibrium becomes unstable.  The destabilization of the equilibrium arises because of the sensitivity of MAC weathering to hydrology, emphasizing a need for greater attention to the interplay of weathering and h
ydroclimate changes in the outer regions of the conventional HZ. Because of limitations in our modeling framework and parameter coverage, our results are not yet sufficient to conclude that the geochemically-consistent HZ is contracted relative to the conventional HZ that only takes into account radiative and thermodynamic constraints, but it does reveal mechanisms where habitability can break down in the outer portions of the conventional HZ. 
\section{Data availability statement}
Isca simulation outputs and Python scripts used to produce the figures in this paper are available for download at\dataset[DOI: 10.5281/zenodo.10995044]{https://doi.org/10.5281/zenodo.10995044} \citep{graham_data_2024}.
\section*{Acknowledgments}
This work was previously published in somewhat modified form as a chapter of RJG's DPhil thesis \cite{graham2022silicate}. We thank Itay Halevy and Vivien Parmentier for reviewing it as a thesis chapter. We also thank two anonymous reviewers that provided useful feedback that significantly improved the manuscript. RJG acknowledges support from the Clarendon Fund and Jesus College, Oxford. This work received support from the UK Science and Technologies Facilities Council Consolidated Grant ST/W000903/1. This AEThER publication is also funded in part by the Alfred P. Sloan Foundation under grant G202114194. Part of this work was also completed with sponsorship by the National Aeronautics and Space Administration (NASA) through a contract with Oak Ridge Associated Universities (ORAU). The views and conclusions contained in this document are those of the authors and should not be interpreted as representing the official policies, either expressed or implied, of the National Aeronautics and Space Administration (NASA) or the U.S. Government. The U.S. Government is authorized to reproduce and distribute reprints for Government purposes notwithstanding any copyright notation herein.
\bibliography{./biblio2.bib}        

\begin{thebibliography}{}
\expandafter\ifx\csname natexlab\endcsname\relax\def\natexlab#1{#1}\fi
\providecommand{\url}[1]{\href{#1}{#1}}

\bibitem[{Abbot(2016)}]{abbot2016analytical}
Abbot, D.~S. 2016, The Astrophysical Journal, 827, 117

\bibitem[{Abbot {et~al.}(2012)Abbot, Cowan, \& Ciesla}]{abbot12-weathering}
Abbot, D.~S., Cowan, N.~B., \& Ciesla, F.~J. 2012, Astrophysical Journal, 756,
  178, {doi:10.1088/0004-637X/756/2/178}

\bibitem[{Allan {et~al.}(2020)Allan, Barlow, Byrne, Cherchi, Douville, Fowler,
  Gan, Pendergrass, Rosenfeld, Swann, {et~al.}}]{allan2020advances}
Allan, R.~P., Barlow, M., Byrne, M.~P., {et~al.} 2020, Annals of the New York
  Academy of Sciences, 1472, 49

\bibitem[{Bandstra \& Brantley(2008)}]{bandstra2008data}
Bandstra, J.~Z., \& Brantley, S.~L. 2008, in Kinetics of Water-Rock Interaction
  (Springer), 211--257

\bibitem[{Baranov {et~al.}(2004)Baranov, Lafferty, \&
  Fraser}]{baranov2004infrared}
Baranov, Y.~I., Lafferty, W.~J., \& Fraser, G.~T. 2004, Journal of molecular
  spectroscopy, 228, 432

\bibitem[{Baum {et~al.}(2022)Baum, Fu, \& Bourguet}]{baum2022sensitive}
Baum, M., Fu, M., \& Bourguet, S. 2022, Geophysical Research Letters,
  e2022GL098843

\bibitem[{Berner(1994)}]{Berner:1994p3295}
Berner, R. 1994, Am J Sci, 294, 56

\bibitem[{Betts \& Miller(1993)}]{betts1993betts}
Betts, A.~K., \& Miller, M.~J. 1993, in The representation of cumulus
  convection in numerical models (Springer), 107--121

\bibitem[{Boer(1993)}]{Boer93}
Boer, G.~J. 1993, Climate Dynamics, 8, 225

\bibitem[{Brady \& Gislason(1997)}]{Brady:1997p3530}
Brady, P., \& Gislason, S. 1997, Geochim Cosmochim Ac, 61, 965

\bibitem[{Brady(1991)}]{brady1991effect}
Brady, P.~V. 1991, Journal of Geophysical Research: Solid Earth, 96, 18101

\bibitem[{Brantley {et~al.}(2008)Brantley, Kubicki, \&
  White}]{brantley2008kinetics}
Brantley, S.~L., Kubicki, J.~D., \& White, A.~F. 2008

\bibitem[{Byrne \& O’Gorman(2015)}]{byrne2015response}
Byrne, M.~P., \& O’Gorman, P.~A. 2015, Journal of Climate, 28, 8078

\bibitem[{Caldeira(1995)}]{CALDEIRA:1995p2285}
Caldeira, K. 1995, Am J Sci, 295, 1077

\bibitem[{Carroll \& Knauss(2005)}]{carroll2005dependence}
Carroll, S.~A., \& Knauss, K.~G. 2005, Chemical Geology, 217, 213

\bibitem[{Catling \& Kasting(2017)}]{catling2017atmospheric}
Catling, D.~C., \& Kasting, J.~F. 2017, Atmospheric evolution on inhabited and
  lifeless worlds (Cambridge University Press)

\bibitem[{Chambers(2020)}]{chambers2020effect}
Chambers, J. 2020, The Astrophysical Journal, 896, 96

\bibitem[{Chen \& Brantley(1998)}]{chen1998diopside}
Chen, Y., \& Brantley, S.~L. 1998, Chemical geology, 147, 233

\bibitem[{Coogan \& Gillis(2020)}]{coogan2020average}
Coogan, L., \& Gillis, K. 2020, Earth and Planetary Science Letters, 536,
  116151

\bibitem[{Coogan \& Dosso(2015)}]{coogan2015alteration}
Coogan, L.~A., \& Dosso, S.~E. 2015, Earth and Planetary Science Letters, 415,
  38

\bibitem[{Coogan \& Gillis(2018)}]{coogan2018temperature}
Coogan, L.~A., \& Gillis, K. 2018, Geochimica et Cosmochimica Acta, 243, 24

\bibitem[{Coogan \& Gillis(2013)}]{coogan2013evidence}
Coogan, L.~A., \& Gillis, K.~M. 2013, Geochemistry, Geophysics, Geosystems, 14,
  1771

\bibitem[{Coy(2022)}]{coy2022diskworld}
Coy, B.~P. 2022, PhD thesis, UNIVERSITY OF CHICAGO

\bibitem[{Cronin(2014)}]{cronin2014choice}
Cronin, T.~W. 2014, Journal of the Atmospheric Sciences, 71, 2994

\bibitem[{Ding \& Wordsworth(2020)}]{ding2020stabilization}
Ding, F., \& Wordsworth, R.~D. 2020, The Astrophysical Journal Letters, 891,
  L18

\bibitem[{Edson {et~al.}(2012)Edson, Kasting, Pollard, Lee, \&
  Bannon}]{edson2012carbonate}
Edson, A.~R., Kasting, J.~F., Pollard, D., Lee, S., \& Bannon, P.~R. 2012,
  Astrobiology, 12, 562

\bibitem[{Edwards \& Slingo(1996)}]{edwards1996studies}
Edwards, J., \& Slingo, A. 1996, Quarterly Journal of the Royal Meteorological
  Society, 122, 689

\bibitem[{Farneti \& Vallis(2013)}]{farneti2013meridional}
Farneti, R., \& Vallis, G.~K. 2013, Journal of climate, 26, 7151

\bibitem[{Feng \& Zhang(2015)}]{feng2015global}
Feng, H., \& Zhang, M. 2015, Scientific reports, 5, 1

\bibitem[{Ferrier \& Kirchner(2008)}]{ferrier2008effects}
Ferrier, K.~L., \& Kirchner, J.~W. 2008, Earth and Planetary Science Letters,
  272, 591

\bibitem[{Foley(2019)}]{foley2019habitability}
Foley, B.~J. 2019, The Astrophysical Journal, 875, 72

\bibitem[{Foley \& Smye(2018)}]{foley2018carbon}
Foley, B.~J., \& Smye, A.~J. 2018, Astrobiology, 18, 873

\bibitem[{Francois \& Walker(1992)}]{francois1992modelling}
Francois, L.~M., \& Walker, J. 1992, American Journal of Science, 292, 81

\bibitem[{Frierson {et~al.}(2006)Frierson, Held, \&
  {Zurita-Gotor}}]{frierson2006gray}
Frierson, D. M.~W., Held, I.~M., \& {Zurita-Gotor}, P. 2006, Journal of the
  Atmospheric Sciences, 63, 2548

\bibitem[{Galili {et~al.}(2019)Galili, Shemesh, Yam, Brailovsky, Sela-Adler,
  Schuster, Collom, Bekker, Planavsky, Macdonald,
  {et~al.}}]{galili2019geologic}
Galili, N., Shemesh, A., Yam, R., {et~al.} 2019, Science, 365, 469

\bibitem[{{Gaudi} {et~al.}(2020){Gaudi}, {Seager}, {Mennesson}, {Kiessling},
  {Warfield}, {Cahoy}, {Clarke}, {Domagal-Goldman}, {Feinberg}, {Guyon},
  {Kasdin}, {Mawet}, {Plavchan}, {Robinson}, {Rogers}, {Scowen}, {Somerville},
  {Stapelfeldt}, {Stark}, {Stern}, {Turnbull}, {Amini}, {Kuan}, {Martin},
  {Morgan}, {Redding}, {Stahl}, {Webb}, {Alvarez-Salazar}, {Arnold}, {Arya},
  {Balasubramanian}, {Baysinger}, {Bell}, {Below}, {Benson}, {Blais}, {Booth},
  {Bourgeois}, {Bradford}, {Brewer}, {Brooks}, {Cady}, {Caldwell}, {Calvet},
  {Carr}, {Chan}, {Cormarkovic}, {Coste}, {Cox}, {Danner}, {Davis}, {Dewell},
  {Dorsett}, {Dunn}, {East}, {Effinger}, {Eng}, {Freebury}, {Garcia}, {Gaskin},
  {Greene}, {Hennessy}, {Hilgemann}, {Hood}, {Holota}, {Howe}, {Huang}, {Hull},
  {Hunt}, {Hurd}, {Johnson}, {Kissil}, {Knight}, {Kolenz}, {Kraus}, {Krist},
  {Li}, {Lisman}, {Mandic}, {Mann}, {Marchen}, {Marrese-Reading}, {McCready},
  {McGown}, {Missun}, {Miyaguchi}, {Moore}, {Nemati}, {Nikzad}, {Nissen},
  {Novicki}, {Perrine}, {Pineda}, {Polanco}, {Putnam}, {Qureshi}, {Richards},
  {Eldorado Riggs}, {Rodgers}, {Rud}, {Saini}, {Scalisi}, {Scharf}, {Schulz},
  {Serabyn}, {Sigrist}, {Sikkia}, {Singleton}, {Shaklan}, {Smith}, {Southerd},
  {Stahl}, {Steeves}, {Sturges}, {Sullivan}, {Tang}, {Taras}, {Tesch},
  {Therrell}, {Tseng}, {Valente}, {Van Buren}, {Villalvazo}, {Warwick}, {Webb},
  {Westerhoff}, {Wofford}, {Wu}, {Woo}, {Wood}, {Ziemer}, {Arney}, {Anderson},
  {Ma{\'\i}z-Apell{\'a}niz}, {Bartlett}, {Belikov}, {Bendek}, {Cenko},
  {Douglas}, {Dulz}, {Evans}, {Faramaz}, {Feng}, {Ferguson}, {Follette},
  {Ford}, {Garc{\'\i}a}, {Geha}, {Gelino}, {G{\"o}tberg}, {Hildebrand t}, {Hu},
  {Jahnke}, {Kennedy}, {Kreidberg}, {Isella}, {Lopez}, {Marchis}, {Macri},
  {Marley}, {Matzko}, {Mazoyer}, {McCandliss}, {Meshkat}, {Mordasini},
  {Morris}, {Nielsen}, {Newman}, {Petigura}, {Postman}, {Reines}, {Roberge},
  {Roederer}, {Ruane}, {Schwieterman}, {Sirbu}, {Spalding}, {Teplitz},
  {Tumlinson}, {Turner}, {Werk}, {Wofford}, {Wyatt}, {Young}, \&
  {Zellem}}]{HABEX_StudyReport2019}
{Gaudi}, B.~S., {Seager}, S., {Mennesson}, B., {et~al.} 2020, arXiv e-prints,
  arXiv:2001.06683

\bibitem[{Ghiggi {et~al.}(2019)Ghiggi, Humphrey, Seneviratne, \&
  Gudmundsson}]{ghiggi2019grun}
Ghiggi, G., Humphrey, V., Seneviratne, S.~I., \& Gudmundsson, L. 2019, Earth
  System Science Data, 11, 1655

\bibitem[{Goldblatt {et~al.}(2021)Goldblatt, McDonald, \&
  McCusker}]{goldblatt2021earth}
Goldblatt, C., McDonald, V.~L., \& McCusker, K.~E. 2021, Nature Geoscience, 1

\bibitem[{Golubev {et~al.}(2005)Golubev, Pokrovsky, \&
  Schott}]{golubev2005experimental}
Golubev, S.~V., Pokrovsky, O.~S., \& Schott, J. 2005, Chemical Geology, 217,
  227

\bibitem[{G{\'o}mez-Leal {et~al.}(2018)G{\'o}mez-Leal, Kaltenegger, Lucarini,
  \& Lunkeit}]{gomez2018climate}
G{\'o}mez-Leal, I., Kaltenegger, L., Lucarini, V., \& Lunkeit, F. 2018, The
  Astrophysical Journal, 869, 129

\bibitem[{Goodwin(2021)}]{goodwin2021probabilistic}
Goodwin, P. 2021, Oxford Open Climate Change, 1, kgab007

\bibitem[{{Gordon} {et~al.}(2017){Gordon}, {Rothman}, {Hill}, {Kochanov},
  {Tan}, {Bernath}, {Birk}, {Boudon}, {Campargue}, {Chance}, {Drouin}, {Flaud},
  {Gamache}, {Hodges}, {Jacquemart}, {Perevalov}, {Perrin}, {Shine}, {Smith},
  {Tennyson}, {Toon}, {Tran}, {Tyuterev}, {Barbe}, {Cs{\'a}sz{\'a}r}, {Devi},
  {Furtenbacher}, {Harrison}, {Hartmann}, {Jolly}, {Johnson}, {Karman},
  {Kleiner}, {Kyuberis}, {Loos}, {Lyulin}, {Massie}, {Mikhailenko},
  {Moazzen-Ahmadi}, {M{\"u}ller}, {Naumenko}, {Nikitin}, {Polyansky}, {Rey},
  {Rotger}, {Sharpe}, {Sung}, {Starikova}, {Tashkun}, {Auwera}, {Wagner},
  {Wilzewski}, {Wcis{\l}o}, {Yu}, \& {Zak}}]{HITRAN2016}
{Gordon}, I.~E., {Rothman}, L.~S., {Hill}, C., {et~al.} 2017, Journal of
  Quantitative Spectroscopy and Radiative Transfer, 203, 3

\bibitem[{Graham(2022)}]{graham2022silicate}
Graham, R. 2022, PhD thesis, University of Oxford

\bibitem[{Graham {et~al.}(2022)Graham, Lichtenberg, \&
  Pierrehumbert}]{graham2022co2}
Graham, R., Lichtenberg, T., \& Pierrehumbert, R.~T. 2022, Journal of
  Geophysical Research: Planets, e2022JE007456

\bibitem[{Graham \& Pierrehumbert(2020)}]{graham2020thermodynamic}
Graham, R., \& Pierrehumbert, R. 2020, Astrophysical Journal, 896

\bibitem[{Graham \& Pierrehumbert(2024)}]{graham_data_2024}
---. 2024, Data and code for "{Carbon} cycle instability for high-{CO2}
  exoplanets: implications for habitability", , , doi:10.5281/zenodo.10995044.
\newblock \url{https://zenodo.org/records/10995044}

\bibitem[{Graham(2021)}]{graham2021high}
Graham, R.~J. 2021, Astrobiology, 21, 1406

\bibitem[{Guimond \& Cowan(2018)}]{guimond2018direct}
Guimond, C.~M., \& Cowan, N.~B. 2018, The Astronomical Journal, 155, 230

\bibitem[{Guimond {et~al.}(2021)Guimond, Noack, Ortenzi, \&
  Sohl}]{guimond2021low}
Guimond, C.~M., Noack, L., Ortenzi, G., \& Sohl, F. 2021, Physics of the Earth
  and Planetary Interiors, 320, 106788

\bibitem[{Guimond {et~al.}(2022)Guimond, Rudge, \& Shorttle}]{guimond2022blue}
Guimond, C.~M., Rudge, J.~F., \& Shorttle, O. 2022, The Planetary Science
  Journal, 3, 66

\bibitem[{Gutowski~Jr {et~al.}(1991)Gutowski~Jr, Gutzler, \&
  Wang}]{gutowski1991surface}
Gutowski~Jr, W.~J., Gutzler, D.~S., \& Wang, W.-C. 1991, Journal of climate, 4,
  121

\bibitem[{Guzewich {et~al.}(2021)Guzewich, Way, Aleinov, Wolf, Del~Genio,
  Wordsworth, \& Tsigaridis}]{guzewich20213d}
Guzewich, S.~D., Way, M.~J., Aleinov, I., {et~al.} 2021, Journal of Geophysical
  Research: Planets, 126, e2021JE006825

\bibitem[{Hakim {et~al.}(2021)Hakim, Bower, Tian, Deitrick, Auclair-Desrotour,
  Kitzmann, Dorn, Mezger, \& Heng}]{hakim2021lithologic}
Hakim, K., Bower, D.~J., Tian, M., {et~al.} 2021, The Planetary Science
  Journal, 2, 49

\bibitem[{Halevy {et~al.}(2009)Halevy, Pierrehumbert, \& Schrag}]{Halevy09}
Halevy, I., Pierrehumbert, R.~T., \& Schrag, D.~P. 2009, Journal of Geophysical
  Research-atmospheres, 114

\bibitem[{Haqq-Misra {et~al.}(2016)Haqq-Misra, Kopparapu, Batalha, Harman, \&
  Kasting}]{haqq2016limit}
Haqq-Misra, J., Kopparapu, R.~K., Batalha, N.~E., Harman, C.~E., \& Kasting,
  J.~F. 2016, The Astrophysical Journal, 827, 120

\bibitem[{Hayworth \& Foley(2020)}]{hayworth2020waterworlds}
Hayworth, B.~P., \& Foley, B.~J. 2020, The Astrophysical Journal Letters, 902,
  L10

\bibitem[{Heimsath {et~al.}(2000)Heimsath, Chappell, Dietrich, Nishiizumi, \&
  Finkel}]{heimsath2000soil}
Heimsath, A.~M., Chappell, J., Dietrich, W.~E., Nishiizumi, K., \& Finkel,
  R.~C. 2000, Geology, 28, 787

\bibitem[{Held \& Soden(2006)}]{held2006robust}
Held, I.~M., \& Soden, B.~J. 2006, Journal of climate, 19, 5686

\bibitem[{Herwartz {et~al.}(2021)Herwartz, Pack, \& Nagel}]{herwartz2021co2}
Herwartz, D., Pack, A., \& Nagel, T.~J. 2021, Proceedings of the National
  Academy of Sciences, 118, e2023617118

\bibitem[{Herweijer {et~al.}(2005)Herweijer, Seager, Winton, \&
  Clement}]{herweijer2005ocean}
Herweijer, C., Seager, R., Winton, M., \& Clement, A. 2005, Tellus A: Dynamic
  Meteorology and Oceanography, 57, 662

\bibitem[{Hilley {et~al.}(2010)Hilley, Chamberlain, Moon, Porder, \&
  Willett}]{hilley2010competition}
Hilley, G., Chamberlain, C., Moon, S., Porder, S., \& Willett, S. 2010, Earth
  and Planetary Science Letters, 293, 191

\bibitem[{Hu \& Yang(2014)}]{hu2014role}
Hu, Y., \& Yang, J. 2014, Proceedings of the National Academy of Sciences, 111,
  629

\bibitem[{Isson \& Planavsky(2018)}]{isson2018reverse}
Isson, T.~T., \& Planavsky, N.~J. 2018, Nature, 560, 471

\bibitem[{Jansen {et~al.}(2019)Jansen, Scharf, Way, \&
  Del~Genio}]{jansen2019climates}
Jansen, T., Scharf, C., Way, M., \& Del~Genio, A. 2019, The Astrophysical
  Journal, 875, 79

\bibitem[{Kadoya {et~al.}(2020)Kadoya, Krissansen-Totton, \&
  Catling}]{kadoya2020probable}
Kadoya, S., Krissansen-Totton, J., \& Catling, D.~C. 2020, Geochemistry,
  Geophysics, Geosystems, 21, e2019GC008734

\bibitem[{Kadoya \& Tajika(2019)}]{kadoya2019outer}
Kadoya, S., \& Tajika, E. 2019, The Astrophysical Journal, 875, 7

\bibitem[{Kalderon-Asael {et~al.}(2021)Kalderon-Asael, Katchinoff, Planavsky,
  Hood, Dellinger, Bellefroid, Jones, Hofmann, Ossa, Macdonald,
  {et~al.}}]{kalderon2021lithium}
Kalderon-Asael, B., Katchinoff, J.~A., Planavsky, N.~J., {et~al.} 2021, Nature,
  595, 394

\bibitem[{Kasting \& Ackerman(1986)}]{KASTING:1986p3082}
Kasting, J.~F., \& Ackerman, T.~P. 1986, Science, 234, 1383

\bibitem[{Kasting \& Harman(2013)}]{kasting2013inner}
Kasting, J.~F., \& Harman, C.~E. 2013, Nature, 504, 221

\bibitem[{Kasting {et~al.}(1993)Kasting, Whitmire, \& Reynolds}]{Kasting93}
Kasting, J.~F., Whitmire, D.~P., \& Reynolds, R.~T. 1993, Icarus, 101, 108

\bibitem[{Kite {et~al.}(2011)Kite, Gaidos, \& Manga}]{Kite:2011}
Kite, E., Gaidos, E., \& Manga, M. 2011, Astrophys Journal, 743,
  {doi:10.1088/0004-637X/743/1/41}

\bibitem[{Kite(2019)}]{kite2019geologic}
Kite, E.~S. 2019, Space Science Reviews, 215, 10

\bibitem[{Kite {et~al.}(2019)Kite, Mayer, Wilson, Davis, Lucas, \& Stucky~de
  Quay}]{kite2019persistence}
Kite, E.~S., Mayer, D.~P., Wilson, S.~A., {et~al.} 2019, Science Advances, 5,
  eaav7710

\bibitem[{Kite {et~al.}(2021)Kite, Steele, Mischna, \&
  Richardson}]{kite2021warm}
Kite, E.~S., Steele, L.~J., Mischna, M.~A., \& Richardson, M.~I. 2021,
  Proceedings of the National Academy of Sciences, 118

\bibitem[{Kitzmann(2017)}]{kitzmann2017clouds}
Kitzmann, D. 2017, Astronomy \& Astrophysics, 600, A111

\bibitem[{Knauss {et~al.}(1993)Knauss, Nguyen, \& Weed}]{knauss1993diopside}
Knauss, K.~G., Nguyen, S.~N., \& Weed, H.~C. 1993, Geochimica et Cosmochimica
  Acta, 57, 285

\bibitem[{Knutti {et~al.}(2017)Knutti, Rugenstein, \&
  Hegerl}]{knutti2017beyond}
Knutti, R., Rugenstein, M.~A., \& Hegerl, G.~C. 2017, Nature Geoscience, 10,
  727

\bibitem[{Kopparapu(2013)}]{kopparapu2013revised}
Kopparapu, R.~K. 2013, The Astrophysical Journal Letters, 767, L8

\bibitem[{Kopparapu {et~al.}(2013)Kopparapu, Ramirez, Kasting, Eymet, Robinson,
  Mahadevan, Terrien, Domagal-Goldman, Meadows, \& Deshpande}]{Kopparapu:2013}
Kopparapu, R.~K., Ramirez, R., Kasting, J.~F., {et~al.} 2013, The Astrophysical
  Journal, 765, 131

\bibitem[{Krissansen-Totton {et~al.}(2018)Krissansen-Totton, Arney, \&
  Catling}]{krissansen2018constraining}
Krissansen-Totton, J., Arney, G.~N., \& Catling, D.~C. 2018, Proceedings of the
  National Academy of Sciences, 115, 4105

\bibitem[{Krissansen-Totton \& Catling(2017)}]{krissansen2017constraining}
Krissansen-Totton, J., \& Catling, D.~C. 2017, Nature communications, 8, 1

\bibitem[{Krissansen-Totton \& Catling(2020)}]{krissansen2020coupled}
---. 2020, Earth and Planetary Science Letters, 537, 116181

\bibitem[{Kump(2018)}]{kump2018prolonged}
Kump, L.~R. 2018, Philosophical Transactions of the Royal Society A:
  Mathematical, Physical and Engineering Sciences, 376, 20170078

\bibitem[{Kundzewicz(2008)}]{kundzewicz2008climate}
Kundzewicz, Z.~W. 2008, Ecohydrology \& Hydrobiology, 8, 195

\bibitem[{Labont{\'e} \& Merlis(2020)}]{labonte2020sensitivity}
Labont{\'e}, M.-P., \& Merlis, T.~M. 2020, The Astrophysical Journal, 896, 31

\bibitem[{{Le Hir} {et~al.}(2009){Le Hir}, Donnadieu, Godderis, Pierrehumbert,
  Halverson, Macouin, Nedelec, \& Ramstein}]{lehir2009snowball}
{Le Hir}, G., Donnadieu, Y., Godderis, Y., {et~al.} 2009, Earth Planet Sc Lett,
  277, 453, {doi:10.1016/j.epsl.2008.11.010}

\bibitem[{Leconte(2018)}]{leconte2018continuous}
Leconte, J. 2018, Nature geoscience, 11, 168

\bibitem[{Li {et~al.}(2006)Li, Scinocca, Lazare, McFarlane, Von~Salzen, \&
  Solheim}]{li2006ocean}
Li, J., Scinocca, J., Lazare, M., {et~al.} 2006, Journal of climate, 19, 6314

\bibitem[{Lowe {et~al.}(2020)Lowe, Ibarra, Drabon, \&
  Chamberlain}]{lowe2020constraints}
Lowe, D.~R., Ibarra, D.~E., Drabon, N., \& Chamberlain, C.~P. 2020, American
  Journal of Science, 320, 790

\bibitem[{Maher \& Chamberlain(2014)}]{maher2014hydrologic}
Maher, K., \& Chamberlain, C. 2014, Science, 343, 1502

\bibitem[{McGrail {et~al.}(2017)McGrail, Schaef, Spane, Cliff, Qafoku, Horner,
  Thompson, Owen, \& Sullivan}]{mcgrail2017field}
McGrail, B.~P., Schaef, H.~T., Spane, F.~A., {et~al.} 2017, Environmental
  Science \& Technology Letters, 4, 6

\bibitem[{McGunnigle {et~al.}(2022)McGunnigle, Cano, Sharp, Muehlenbachs, Cole,
  Hardman, Stachel, \& Pearson}]{mcgunnigle2022triple}
McGunnigle, J., Cano, E., Sharp, Z., {et~al.} 2022, Geology

\bibitem[{Menou(2015)}]{Menou2015}
Menou, K. 2015, Earth and Planetary Science Letters, 429, 20

\bibitem[{{Mlawer} {et~al.}(2012){Mlawer}, {Payne}, {Moncet}, {Delamere},
  {Alvarado}, \& {Tobin}}]{mlawer2012development}
{Mlawer}, E.~J., {Payne}, V.~H., {Moncet}, J.~L., {et~al.} 2012, Philos. Trans.
  Royal Soc. A, 370, 2520

\bibitem[{Montgomery \& Brandon(2002)}]{montgomery2002topographic}
Montgomery, D.~R., \& Brandon, M.~T. 2002, Earth and Planetary Science Letters,
  201, 481

\bibitem[{Myhre {et~al.}(2017)Myhre, Forster, Samset, Hodnebrog, Sillmann,
  Aalbergsj{\o}, Andrews, Boucher, Faluvegi, Fl{\"a}schner,
  {et~al.}}]{myhre2017pdrmip}
Myhre, G., Forster, P., Samset, B., {et~al.} 2017, Bulletin of the American
  Meteorological Society, 98, 1185

\bibitem[{Myhre {et~al.}(2018)Myhre, Samset, Hodnebrog, Andrews, Boucher,
  Faluvegi, Fl{\"a}schner, Forster, Kasoar, Kharin,
  {et~al.}}]{myhre2018sensible}
Myhre, G., Samset, B.~H., Hodnebrog, {\O}., {et~al.} 2018, Nature
  Communications, 9, 1922

\bibitem[{Noack {et~al.}(2017)Noack, Rivoldini, \&
  Van~Hoolst}]{noack2017volcanism}
Noack, L., Rivoldini, A., \& Van~Hoolst, T. 2017, Physics of the Earth and
  Planetary Interiors, 269, 40

\bibitem[{Oelkers \& Schott(2001)}]{oelkers2001experimental}
Oelkers, E.~H., \& Schott, J. 2001, Geochimica et Cosmochimica Acta, 65, 1219

\bibitem[{Owen {et~al.}(2011)Owen, Amundson, Dietrich, Nishiizumi, Sutter, \&
  Chong}]{owen2011sensitivity}
Owen, J.~J., Amundson, R., Dietrich, W.~E., {et~al.} 2011, Earth Surface
  Processes and Landforms, 36, 117

\bibitem[{Oxburgh {et~al.}(1994)Oxburgh, Drever, \& Sun}]{oxburgh1994mechanism}
Oxburgh, R., Drever, J.~I., \& Sun, Y.-T. 1994, Geochimica et Cosmochimica
  Acta, 58, 661

\bibitem[{O’Gorman {et~al.}(2012)O’Gorman, Allan, Byrne, \&
  Previdi}]{o2012energetic}
O’Gorman, P.~A., Allan, R.~P., Byrne, M.~P., \& Previdi, M. 2012, Surveys in
  geophysics, 33, 585

\bibitem[{O’Gorman \& Schneider(2008)}]{o2008hydrological}
O’Gorman, P.~A., \& Schneider, T. 2008, Journal of Climate, 21, 3815

\bibitem[{Palandri \& Kharaka(2004)}]{palandri2004compilation}
Palandri, J.~L., \& Kharaka, Y.~K. 2004, A compilation of rate parameters of
  water-mineral interaction kinetics for application to geochemical modeling,
  Tech. rep., Geological Survey Menlo Park CA

\bibitem[{Paradise \& Menou(2017)}]{paradise2017}
Paradise, A., \& Menou, K. 2017, The Astrophysical Journal, 848, 1

\bibitem[{Paradise {et~al.}(2019)Paradise, Menou, Valencia, \&
  Lee}]{paradisehabitable}
Paradise, A., Menou, K., Valencia, D., \& Lee, C. 2019, Journal of Geophysical
  Research: Planets

\bibitem[{Penman {et~al.}(2020)Penman, Rugenstein, Ibarra, \&
  Winnick}]{penman2020silicate}
Penman, D.~E., Rugenstein, J. K.~C., Ibarra, D.~E., \& Winnick, M.~J. 2020,
  Earth-Science Reviews, 103298

\bibitem[{Penn \& Vallis(2018)}]{penn2018atmospheric}
Penn, J., \& Vallis, G.~K. 2018, The Astrophysical Journal, 868, 147

\bibitem[{Perrin \& Hartmann(1989)}]{perrin1989temperature}
Perrin, M., \& Hartmann, J. 1989, Journal of Quantitative Spectroscopy and
  Radiative Transfer, 42, 311

\bibitem[{Pierrehumbert(1999)}]{pierrehumbert1999subtropical}
Pierrehumbert, R.~T. 1999, GEOPHYSICAL MONOGRAPH-AMERICAN GEOPHYSICAL UNION,
  112, 339

\bibitem[{Pierrehumbert(2002)}]{Pierrehumbert-2002:hydrologic}
---. 2002, Nature, 419, 191

\bibitem[{Pierrehumbert(2010)}]{Pierrehumbert:2010-book}
---. 2010, Principles of Planetary Climate (Cambridge University Press)

\bibitem[{Pierrehumbert {et~al.}(2011)Pierrehumbert, Abbot, Voigt, \&
  Koll}]{Pierrehumbert-et-al-2010:neoprot}
Pierrehumbert, R.~T., Abbot, D.~S., Voigt, A., \& Koll, D. 2011, Ann. Rev. of
  Earth and Planet. Sci., 39, 417, {DOI:10.1146/annurev-earth-040809-152447}

\bibitem[{{Quanz} {et~al.}(2021){Quanz}, {Ottiger}, {Fontanet}, {Kammerer},
  {Menti}, {Dannert}, {Gheorghe}, {Absil}, {Airapetian}, {Alei}, {Allart},
  {Angerhausen}, {Blumenthal}, {Buchhave}, {Cabrera},
  {Carri{\'o}n-Gonz{\'a}lez}, {Chauvin}, {Danchi}, {Dandumont}, {Defr{\`e}re},
  {Dorn}, {Ehrenreich}, {Ertel}, {Fridlund}, {Garc{\'\i}a Mu{\~n}oz},
  {Gasc{\'o}n}, {Girard}, {Glauser}, {Grenfell}, {Guidi}, {Hagelberg},
  {Helled}, {Ireland}, {Kopparapu}, {Korth}, {Kozakis}, {Kraus}, {L{\'e}ger},
  {Leedj{\"a}rv}, {Lichtenberg}, {Lillo-Box}, {Linz}, {Liseau}, {Loicq},
  {Mahendra}, {Malbet}, {Mathew}, {Mennesson}, {Meyer}, {Mishra},
  {Molaverdikhani}, {Noack}, {Oza}, {Pall{\'e}}, {Parviainen}, {Quirrenbach},
  {Rauer}, {Ribas}, {Rice}, {Romagnolo}, {Rugheimer}, {Schwieterman},
  {Serabyn}, {Sharma}, {Stassun}, {Szul{\'a}gyi}, {Wang}, {Wunderlich},
  {Wyatt}, \& {the LIFE collaboration}}]{LIFE2021a}
{Quanz}, S.~P., {Ottiger}, M., {Fontanet}, E., {et~al.} 2021, arXiv e-prints,
  arXiv:2101.07500

\bibitem[{Ramirez {et~al.}(2014)Ramirez, Kopparapu, Lindner, \&
  Kasting}]{ramirez2014can}
Ramirez, R.~M., Kopparapu, R.~K., Lindner, V., \& Kasting, J.~F. 2014,
  Astrobiology, 14, 714

\bibitem[{Rimstidt {et~al.}(2012)Rimstidt, Brantley, \&
  Olsen}]{rimstidt2012systematic}
Rimstidt, J.~D., Brantley, S.~L., \& Olsen, A.~A. 2012, Geochimica et
  Cosmochimica Acta, 99, 159

\bibitem[{Roe(2005)}]{roe2005orographic}
Roe, G.~H. 2005, Annual Review of earth and planetary sciences, 33, 645

\bibitem[{Romps(2020)}]{romps2020climate}
Romps, D.~M. 2020, Journal of Climate, 33, 3413

\bibitem[{Rose(2015)}]{rose2015stable}
Rose, B.~E. 2015, Journal of Geophysical Research: Atmospheres, 120, 1404

\bibitem[{Russell {et~al.}(2013)Russell, Lacis, Rind, Colose, \&
  Opstbaum}]{russell2013fast}
Russell, G.~L., Lacis, A.~A., Rind, D.~H., Colose, C., \& Opstbaum, R.~F. 2013,
  Geophysical Research Letters, 40, 5787

\bibitem[{Schneider {et~al.}(2019)Schneider, Kaul, \&
  Pressel}]{schneider2019possible}
Schneider, T., Kaul, C.~M., \& Pressel, K.~G. 2019, Nature Geoscience, 12, 163

\bibitem[{Schott \& Berner(1985)}]{schott1985dissolution}
Schott, J., \& Berner, R.~A. 1985, in The chemistry of weathering (Springer),
  35--53

\bibitem[{Sellers(1969)}]{Sellers-1969:ae}
Sellers, W.~D. 1969, J. Appl. Meteor., 8, 392

\bibitem[{Shibuya \& Takai(2022)}]{shibuya2022liquid}
Shibuya, T., \& Takai, K. 2022, Progress in Earth and Planetary Science, 9, 1

\bibitem[{Siever(1992)}]{siever1992silica}
Siever, R. 1992, Geochimica et Cosmochimica Acta, 56, 3265

\bibitem[{Siler {et~al.}(2019)Siler, Roe, Armour, \&
  Feldl}]{siler2019revisiting}
Siler, N., Roe, G.~H., Armour, K.~C., \& Feldl, N. 2019, Climate Dynamics, 52,
  3983

\bibitem[{Solomatova \& Caracas(2021)}]{solomatova2021genesis}
Solomatova, N.~V., \& Caracas, R. 2021, Science advances, 7, eabj0406

\bibitem[{Stephens {et~al.}(2015)Stephens, O'Brien, Webster, Pilewski, Kato, \&
  Li}]{stephens2015albedo}
Stephens, G.~L., O'Brien, D., Webster, P.~J., {et~al.} 2015, Reviews of
  geophysics, 53, 141

\bibitem[{Strogatz(1994)}]{Strogatz-1994:nonlinear}
Strogatz, S. 1994, Nonlinear dynamics and chaos (Westview Press)

\bibitem[{Team {et~al.}(2019)}]{luvoir2019luvoir}
Team, L., {et~al.} 2019, arXiv preprint arXiv:1912.06219

\bibitem[{Thomson \& Vallis(2019{\natexlab{a}})}]{thomson2019effects}
Thomson, S.~I., \& Vallis, G.~K. 2019{\natexlab{a}}, Quarterly Journal of the
  Royal Meteorological Society, 145, 2627

\bibitem[{Thomson \& Vallis(2019{\natexlab{b}})}]{thomson2019hierarchical}
---. 2019{\natexlab{b}}, Atmosphere, 10, 803

\bibitem[{Trenberth \& Caron(2001)}]{trenberth2001estimates}
Trenberth, K.~E., \& Caron, J.~M. 2001, Journal of Climate, 14, 3433

\bibitem[{Turbet {et~al.}(2017)Turbet, Forget, Leconte, Charnay, \&
  Tobie}]{turbet2017co}
Turbet, M., Forget, F., Leconte, J., Charnay, B., \& Tobie, G. 2017, Earth and
  Planetary Science Letters, 11, 11

\bibitem[{Vallis {et~al.}(2018)Vallis, Colyer, Geen, Gerber, Jucker, Maher,
  Paterson, Pietschnig, Penn, \& Thomson}]{vallis2018isca}
Vallis, G.~K., Colyer, G., Geen, R., {et~al.} 2018, Geoscientific Model
  Development

\bibitem[{Vecchi {et~al.}(2006)Vecchi, Soden, Wittenberg, Held, Leetmaa, \&
  Harrison}]{vecchi2006weakening}
Vecchi, G.~A., Soden, B.~J., Wittenberg, A.~T., {et~al.} 2006, Nature, 441, 73

\bibitem[{V{\'e}rard \& Veizer(2019)}]{verard2019plate}
V{\'e}rard, C., \& Veizer, J. 2019, Geology, 47, 881

\bibitem[{Von~Paris {et~al.}(2013)Von~Paris, Selsis, Kitzmann, \&
  Rauer}]{von2013dependence}
Von~Paris, P., Selsis, F., Kitzmann, D., \& Rauer, H. 2013, Astrobiology, 13,
  899

\bibitem[{Walker {et~al.}(1981)Walker, Hays, \&
  Kasting}]{Walker-Hays-Kasting-1981:negative}
Walker, J.~C.~G., Hays, P.~B., \& Kasting, J.~F. 1981, Journal of Geophysical
  Research, 86, 9776

\bibitem[{Walters {et~al.}(2019)Walters, Baran, Boutle, Brooks, Earnshaw,
  Edwards, Furtado, Hill, Lock, Manners, {et~al.}}]{walters2019met}
Walters, D., Baran, A.~J., Boutle, I., {et~al.} 2019, Geoscientific Model
  Development, 12, 1909

\bibitem[{Way {et~al.}(2017)Way, Aleinov, Amundsen, Chandler, Clune, Del~Genio,
  Fujii, Kelley, Kiang, Sohl, \& Tsigaridis}]{way2017resolving}
Way, M.~J., Aleinov, I., Amundsen, D.~S., {et~al.} 2017, The Astrophysical
  Journal Supplement Series, 231, 12

\bibitem[{Weissbart \& Rimstidt(2000)}]{weissbart2000wollastonite}
Weissbart, E.~J., \& Rimstidt, J.~D. 2000, Geochimica et Cosmochimica Acta, 64,
  4007

\bibitem[{Welch \& Ullman(1996)}]{welch1996feldspar}
Welch, S., \& Ullman, W. 1996, Geochimica et Cosmochimica Acta, 60, 2939

\bibitem[{West(2012)}]{west2012thickness}
West, A.~J. 2012, Geology, 40, 811

\bibitem[{Winnick \& Maher(2018)}]{winnick2018relationships}
Winnick, M.~J., \& Maher, K. 2018, Earth and Planetary Science Letters, 485,
  111

\bibitem[{Wolf {et~al.}(2018)Wolf, Haqq-Misra, \& Toon}]{wolf2018evaluating}
Wolf, E., Haqq-Misra, J., \& Toon, O. 2018, Journal of Geophysical Research:
  Atmospheres, 123, 11

\bibitem[{Wordsworth {et~al.}(2010)Wordsworth, Forget, \&
  Eymet}]{wordsworth2010infrared}
Wordsworth, R., Forget, F., \& Eymet, V. 2010, Icarus, 210, 992

\bibitem[{Wordsworth \& Pierrehumbert(2013)}]{wordsworth2013water}
Wordsworth, R.~D., \& Pierrehumbert, R.~T. 2013, The Astrophysical Journal,
  778, 154

\bibitem[{Xiong {et~al.}(2022)Xiong, Yang, \& Liu}]{xiong2022smaller}
Xiong, J., Yang, J., \& Liu, J. 2022, Geophysical Research Letters, 49,
  e2022GL099599

\bibitem[{Yang {et~al.}(2019{\natexlab{a}})Yang, Abbot, Koll, Hu, \&
  Showman}]{yang2019ocean}
Yang, J., Abbot, D.~S., Koll, D.~D., Hu, Y., \& Showman, A.~P.
  2019{\natexlab{a}}, The Astrophysical Journal, 871, 29

\bibitem[{Yang {et~al.}(2014)Yang, Bou\'e, Fabrycky, \& Abbot}]{yang2014}
Yang, J., Bou\'e, G., Fabrycky, D.~C., \& Abbot, D.~S. 2014, Astrophysical
  Journal Letters, 787, {L2, doi:10.1088/2041-8205/787/1/L2}

\bibitem[{Yang {et~al.}(2013)Yang, Cowan, \& Abbot}]{yang2013}
Yang, J., Cowan, N.~B., \& Abbot, D.~S. 2013, Astrophysical Journal Letters,
  771, {L45, DOI:10.1088/2041-8205/771/2/L45}

\bibitem[{Yang {et~al.}(2019{\natexlab{b}})Yang, Leconte, Wolf, Merlis, Koll,
  Forget, \& Abbot}]{yang2019simulations}
Yang, J., Leconte, J., Wolf, E.~T., {et~al.} 2019{\natexlab{b}}, The
  Astrophysical Journal, 875, 46

\end{thebibliography}
\bibliographystyle{aasjournal.bst} 

\end{document}